\documentclass[reprint,nofootinbib,amsmath,amssymb,aps,eqsecnum,showpacs,twocolumn,superscriptaddress,prb,amsart]{revtex4-2}
\usepackage{graphicx}
\usepackage[nointegrals]{wasysym} 
\usepackage[export]{adjustbox}
\usepackage{amsmath,amsfonts,amssymb,latexsym}
\usepackage{hhline}
\usepackage{bm}
\usepackage{verbatim}
\usepackage{enumitem}
\usepackage{mathrsfs}
\usepackage{slashed}
\usepackage{empheq}
\newcommand{\proj}[1]{|#1\rangle\langle #1|}

\newcommand{\Tr}{\text{Tr}}

\newcommand{\p}{\partial}

\newcommand{\lan}{\langle}
\newcommand{\ran}{\rangle}

\newcommand{\unit}{\mathbf{1}}

\newcommand{\da}{{\dagger}}

\newcommand{\ra}{\rightarrow}

\renewcommand{\(}{\left(}
\renewcommand{\)}{\right)}

\newcommand{\tp}{\otimes}

\newcommand{\twp}{{2\pi}}

\newcommand\bpm            {\begin{pmatrix}}
\newcommand\epm           {\end{pmatrix}}

\def\app#1#2{	\mathrel{		\setbox0=\hbox{$#1\sim$}		\setbox2=\hbox{			\rlap{\hbox{$#1\propto$}}			\lower1.1\ht0\box0		}		\raise0.25\ht2\box2	}}

\newcommand{\vp}{\varphi}

\newcommand{\inv}{^{-1}}

\newcommand{\ope}\odot

\usepackage{manfnt}

\newcommand{\bi}{\begin{itemize}}
	\newcommand{\ei}{\end{itemize}}

\newcommand{\igptc}[1]{\vcenter{\hbox{\includegraphics[width=.3\textwidth]{#1}}}}
\usepackage{amsthm}

\newtheorem{theorem}{Theorem}

\theoremstyle{definition}
\newtheorem{definition}{Definition}
\theoremstyle{definition}

\newcommand\bd            {\begin{definition}}
	\newcommand\ed            {\end{definition}}
\newcommand\bt            {\begin{theorem}}
	\newcommand\et            {\end{theorem}}
\newcommand\be            {\begin{equation}}
	\newcommand\ee            {\end{equation}}
\newcommand\ba            {\begin{aligned}}
	\newcommand\ea            {\end{aligned}}
\newcommand\bea{\begin{equation}\begin{aligned}}
		\newcommand\eea{\end{aligned}\end{equation}}

\usepackage{subcaption}
\usepackage[usenames,dvipsnames]{xcolor}

\usepackage{hyperref} 
\hypersetup{final}
\hypersetup{colorlinks, citecolor=red, linkcolor=red, urlcolor=red} 

\newcommand{\sss}{\subsubsection}
\renewcommand{\ss}{\subsection}

\renewcommand{\a}{\alpha}
\renewcommand{\b}{\beta}

\newcommand{\g}{\gamma}

\renewcommand{\o}{\omega}

\newcommand{\bfa}{\mathbf{a}}

\newcommand{\bfg}{\mathbf{g}}

\newcommand{\zn}{\mathbb{Z}_N}

\newcommand{\zz}{\mathbb{Z}}

\newcommand{\mcu}{\mathcal{U}}
\newcommand{\mce}{\mathcal{E}}

\newcommand{\mcl}{\mathcal{L}}

\newcommand{\mcs}{\mathcal{S}}

\newcommand{\mcr}{\mathcal{R}}
\newcommand{\mcq}{\mathcal{Q}}

\usepackage[mathscr]{eucal}

\usepackage{braket}

\renewcommand{\k}{\ket}

\usepackage{dcolumn}
\usepackage{pifont}

\usepackage{tikz}
\usepackage{ifthen}
\usepackage{framed,xcolor}
\definecolor{shadecolor}{gray}{0.9}

\usepackage{ulem}

\begin{document}
\hfill MIT-CTP/5614
\title{Topological quantum chains protected by dipolar and \\ other modulated symmetries}
 	
\author{Jung Hoon \surname{Han}}
	\email[Electronic address:$~~$]{hanjemme@gmail.com}
	\affiliation{Department of Physics, Sungkyunkwan University, Suwon 16419, South Korea}

\author{Ethan Lake}
\email[Electronic address:$~~$]{elake@mit.edu}
\affiliation{Department of Physics, Massachusetts Institute of Technology, Cambridge, Massachusetts 02139, USA}
\affiliation{Department of Physics, University of California, Berkeley, CA 94720, USA}

\author{Ho Tat Lam}
\email[Electronic address:$~~$]{htlam@mit.edu}
\affiliation{Department of Physics, Massachusetts Institute of Technology, Cambridge, Massachusetts 02139, USA}

\author{Ruben Verresen}
\email[Electronic address:$~~$]{rubenverresen@g.harvard.edu}
\affiliation{Department of Physics, Harvard University, Cambridge, MA 02138, USA}
\affiliation{Department of Physics, Massachusetts Institute of Technology, Cambridge, Massachusetts 02139, USA}

\author{Yizhi You}
\email[Electronic address:$~~$]{y.you@northeastern.com}
\affiliation{Department of Physics, Northeastern University, 360 Huntington Ave, Boston, MA 02115, USA}
\date{\today}

\begin{abstract}
We investigate the physics of one-dimensional symmetry protected topological (SPT) phases protected by symmetries whose symmetry generators exhibit spatial modulation. 
We focus in particular on phases protected by symmetries with linear (i.e., dipolar), quadratic and exponential modulations.
We present a simple recipe for constructing modulated SPT models by generalizing the concept of decorated domain walls to spatially modulated symmetry defects, and develop several tools for characterizing and classifying modulated SPT phases. 
A salient feature of modulated symmetries is that they are generically only present for open chains, and are broken upon the imposition of periodic boundary conditions. Nevertheless, we show that SPT order is present even with periodic boundary conditions, a phenomenon we understand within the context of an object we dub a ``bundle symmetry''. In addition, we show that modulated SPT phases can avoid a certain no-go theorem, leading to an unusual algebraic structure in their matrix product state descriptions.
\end{abstract}

\maketitle

\tableofcontents

\section{Introduction}

Symmetries and patterns of quantum entanglement are two of the major features that characterize different quantum phases of matter. This is most succinctly illustrated in the case of symmetry protected topological (SPT) phases, where global symmetries can stabilize certain patterns of short-range entanglement. 
The zoology of SPT phases protected by different types of symmetries has been carefully categorized over the years~\cite{Su1979-rl,Haldane1983-ya,Affleck1987-wn,Schuch2011-jx,pollmann10,Turner2011-zi,fidkowski2011topological,Chen2011-et,son12,Chen2011-kz,Pollmann2012-lv,Chen2012-oa,Levin2012-dv,Qi13,Vishwanath2013-pb,Yao2013-vc,mesaros13,else2014classifying,Gu2014-lj,chen14,Kapustin:2014tfa,Kapustin:2014dxa,senthil15,Gaiotto2016-ba,Freed:2016rqq,dhlee17,verresen2017,thorngren18,cheng-PRB18,stephen19,cirac19}. In particular, \cite{stephen19,stephen2022universal} have analyzed SPT phases protected by $L$-cycle symmetries in which the protecting symmetry operators periodically vary over $L$ sites \cite{cirac19}.

In another vein, it has been understood that multipole conservation prohibits the free motion of charge and limits the overall phase space of the dynamics, leading to new types of quantum ground states~\cite{pielawa11,sachdev02,pretko17,pretko18,gromov20,seiberg22a,glorioso2023goldstone,lake1,lake2,feldmeier,lake2023non,lake3,seidel05,you20,oh22b,huang2023chern,seiberg23} and unusual dynamical properties, including robust ergodicity breaking~\cite{sala2020ergodicity,khemani2020localization} and anomalous hydrodynamics~\cite{feldmeier20,nandkishore21,han2023scaling,ogunnaike2023unifying,morningstar2023hydrodynamics,gliozzi2023hierarchical,glorioso22,sala2022dynamics,feldmeier20,nandkishore21,Rakovszky20}. These systems have also begun to be studied experimentally, where dipole symmetry---the simplest type of multipolar conservation law---arises in experiments involving ultracold atoms paced in strongly tilted optical lattices~\cite{bakr20,aidelsburger21,kohlert2021experimental,weitenberg22}. 

These symmetries arise in systems that conserve both a global charge and various multipole moments thereof. 
This work is devoted to developing an understanding of short-range entangled phases of matter protected by multipole and exponential symmetries. Much of our intuition in this regard comes from studying an exactly solvable model we introduce, where $\mathbb{Z}_N$ monopole and $\mathbb{Z}_N$ dipole symmetries lead to a new type of SPT order dubbed the dipolar SPT. Despite this model bearing some resemblance to the ordinary $\zn$ cluster state~\cite{Briegel01,son12,geraedts2014exact,Santos15}, we show that there is no unitary which maps between the two models and their protecting symmetry groups, establishing that the dipolar SPT model is new. 

The dipolar SPT phase is in fact only one example in a much broader class of {\it modulated} SPT phases, several examples of which we introduce and study in detail. These phases are protected by global symmetries whose generators are modulated in space, the unusual physics of which is still under active investigation~\cite{sala2022dynamics,gromov2019towards,sala2023exotic}. The trick that allows us to do this is the decorated domain wall scheme \cite{chen2014symmetry} which we exploit to construct exactly solvable SPT models protected by quadrupolar and exponential charge symmetries~\cite{watanabe23,watanabe2022ground,delfino23}. Although formally the symmetry operators protecting this kind of SPT phase can be cyclic in space \cite{stephen19,stephen2022universal} (`$L$-cyclic' according to \cite{cirac19}), we show that cyclicity is not a necessary ingredient in defining the modulated SPT order.

An interesting feature of modulated symmetries is that their symmetry generators in general may not be consistently defined in systems with periodic boundary conditions. This fact raises the question of whether or not their associated symmetry-protected phases are robustly defined in such situations. We show that despite the absence of a global symmetry operator, such SPT phases are still well-defined. We do this by introducing the notion of a {\it bundle symmetry}, a concept more general than a conventional global symmetry. The modulated SPT phases we study in this work are consequently protected by bundle symmetries, rather than the usual global symmetries. The bundle character of the protecting symmetry has not appeared in previous works on SPT classification, but marks the key characteristic of the modulated SPT. 

The paper is organized as follows. In Sec. \ref{sec:2} we review the standard $\zn$ cluster state SPT in one dimension. In Sec. \ref{sec:3} we introduce the dipolar SPT through an exactly solvable model and its ground state exact wave function, and construct its fractionalized edge modes, string operators, and MPS representation. The stability of the dipolar SPT phase under the symmetry-preserving perturbations is checked numerically. The inequivalence of the dipolar SPT to the usual $\zn$ cluster state is carefully discussed. In Sec. \ref{sec:quad-spt} and Sec. \ref{sec:6}, we extend our analysis of dipole-protected SPT phases to phases protected by quadrupolar symmetry and a type of exponentially modulated symmetry, respectively. We end with a summary and perspective in Sec. \ref{sec:7}. 

\section{Review of conventional $\mathbb{Z}_N \times \mathbb{Z}_N$ SPT phases}
\label{sec:2}

In this section, we review the basic properties of one-dimensional SPT phases protected by a conventional onsite $\mathbb{Z}_N \times \mathbb{Z}_N$ symmetry.  This review will serve as preparation for discussing the charge multipole-conserving SPT models to follow; readers already familiar with the details may skip to Sec.~\ref{sec:3}. 

\subsection{The $\mathbb{Z}_N$ cluster model} 

The $\mathbb{Z}_N$ cluster model in 1D is a well-known example of an SPT phase with $\mathbb{Z}_N \times \mathbb{Z}_N$ global symmetry---originally introduced for $N=2$~\cite{Briegel01,son12,chen2014symmetry} and later generalized to arbitrary $N$ \cite{geraedts2014exact,Santos15}. We consider a chain with an $N$-dimensional onsite Hilbert space, whose basis vectors we write as $|g\rangle$, $g \in \mathbb{Z}_N$. 
We will write the Hamiltonian $H_C$ of the cluster state as 
\bea  \label{eq:cluster-H}
H_C & = - \sum_j (b_{2j-1} + b_{2j} + h.c. )  \\ 
b_{2j-1} & = Z^\dag_{2j-2} X_{2j-1} Z_{2j} \\ 
b_{2j} & = Z_{2j-1} X_{2j} Z^\dag_{2j+1} 
\eea 
where we have defined the generalized Pauli operators $X \equiv \sum_g |g+1\ran\lan g|$ (with $g+1$ taken mod $N$) and $Z\equiv \sum_g \o^g |g\ran\lan g|$, with $\o \equiv e^{\twp i /N}$.  
Note that the $b_j$'s are mutually commuting stabilizers. This property makes analyzing the physics of $H_C$ extremely simple.

The model possesses two $\mathbb{Z}_N$ global symmetries $Q_1,Q_2$ supported on even and odd sites of the chain, respectively:  
\begin{align}
Q_{1} = \prod_j X_{2j-1} ,  ~~~ Q_2  = \prod_j X_{2j} ,  \label{eq:charge-symmetry}
\end{align}
with $Q_1^N = Q_2^N = 1$. We refer to these symmetries as {\it monopolar} symmetries, to be distinguished from dipolar and other modulated symmetries introduced in later sections. Monopolar symmetries are distinguished by being generated by ``uniform'' operators, which act the same way in each 2-site unit cell of the lattice.  
This means that translation through two lattice sites --- $i.e.$ one unit cell --- commutes with both $Q_1, Q_2$. 

The ground state $|\Psi_C\ran$ can be obtained by finding a wavefunction that is stabilized by each of the $a_j$. 
This can be done by writing 
\begin{align}
|\Psi_C \rangle \propto \sum_{\mathbf{g} \in (\mathbb{Z}_N )^{\otimes L} } \omega^{\sum_j g_{2j} ( g_{2j-1}  - g_{2j+1} ) } 
|\mathbf{g}\rangle  \label{eq:GS-cluster-H}
\end{align}
where $|\mathbf{g}\rangle \equiv |g_1 , \cdots, g_L \rangle$ ($g_i \in \mathbb{Z}_N$) for a lattice of length $L$.
Defining the state $\k+\equiv \frac{1}{\sqrt{N}}\sum_{g} |g\rangle$, we can write down a unitary which prepares $|\Psi_C\ran$ from a product state as $|\Psi_C \rangle = {\cal U}_C \k+^{\tp L}$, where
\begin{equation}
{\cal U}_C = \prod_{n=1}^{L/2} CZ_{2n-1,2n}  \; \times \; CZ_{2n,2n+1}^\dagger.
\end{equation}
In this notation, which will be employed throughout this work, $C \mathcal O_{mn}$ denotes a conditional two-qudit unitary gate which acts as $\mathcal O^g$ on qudit $n$ if qudit $m$ is in the state $\ket{g}$. Explicitly,
\begin{equation}
C \mathcal O_{mn} = \frac{1}{N} \sum_{\alpha=1}^N \sum_{\beta = 1}^N \omega^{-\alpha\beta} Z_m^\alpha \mathcal O_n^\beta , \label{eq:cond}
\end{equation}
which implies $CZ_{mn}\ket{g_m g_n} = \omega^{g_m g_n}\ket{g_m g_n}$. 

To check that $\mcu_C$ as defined above indeed produces $|\Psi_C\ran$, one can employ the following readily-verified transformations:
\begin{align}
    {\cal U}^\dag_C X_{2j-1} {\cal U}_C & = Z_{2j-2} X_{2j-1} Z^\dag_{2j} ,   \nonumber \\   
    {\cal U}^\dag_C X_{2j} {\cal U}_C & = Z_{2j-1}^\dag X_{2j} Z_{2j+1} .
    \label{eq:transformed-C-terms}
\end{align}
Under this unitary transformation, the cluster Hamiltonian on a periodic chain of even-length $L$ changes to
\begin{align}
{\cal U}^\dag_C H_C {\cal U}_C = - \sum_{j=1}^{L} ( X_j + X_j^\dag )  . 
\label{eq:transformed-C} \end{align}
The state $\k+^{\tp L}$ is obviously the ground state of $X_j + X^\dag_j$ for all $j$, hence $|\Psi_C \rangle = {\cal U}_C  \k+^{\tp L}$ is the ground state of $H_C$. 

\subsection{String order parameters, edge modes and entanglement spectrum \label{subsec:clusterstring}}

String-order operators \cite{denNijs89,sop2012} characterizing the SPT nature of the $\mathbb{Z}_N$ cluster model can be constructed as a product of stabilizers of the cluster Hamiltonian over odd and even sites, respectively:
\begin{align} 
{\cal S}_{Q_1} &= b_{2j+1} b_{2j+3} \cdots b_{2j+2m-1} \nonumber \\ & = Z^{\dag}_{2j} \left( \prod^m_{n=1}   X_{2j+2n-1} \right)  Z_{2j+2m}\nonumber\\
{\cal S}_{Q_2} & = b_{2j} b_{2j+2} \cdots b_{2j+2m-2}   \nonumber \\
&=Z_{2j-1} \left( \prod_{n=1}^m X_{2j+2n-2} \right) Z_{2j+2m-1}^{\dagger} . 
\label{eq:str-for-cluster}
\end{align}
The subscripts on ${\cal S}$ indicate which of the global symmetries is associated with the particular string order parameter. One can view them as measuring the ``monopole'' correlations ($Z^\dag_{2j} Z_{2j+2m}$ or $Z_{2j-1} Z^\dag_{2j+2m-1}$) intertwined with the string order given by the product of $X$ between the two monopoles. Note that the string operator constructed from the product of odd-site stabilizers has monopoles at the even sites ($Z^\dag_{2j}$ and $Z_{2j+2m}$), and vice versa. The expectation values of both string order parameters are strictly equal to one for the cluster ground states \eqref{eq:GS-cluster-H}, and are generically\footnote{Any particular choice of string order can vanish in a measure zero case; however, for any point in the SPT phase there always exists a choice of charged endpoint operator such that the string order is nonzero \cite{sop2012}.} nonzero for ground states of models related to $H_C$ by symmetry-preserving perturbations which do not close the bulk gap.

For an open chain of even length $L$, the unitary transformation $\mathcal{U}_C$ becomes
\begin{align}
\mathcal{U}_C =\prod_{n=1}^{L/2} CZ_{2n-1,2n}  \; \times \; \prod_{n=1}^{L/2-1} CZ_{2n,2n+1}^\dagger.
\label{eq:U_M} \end{align}
(There is a similar unitary for an open chain of odd length.)
This unitary transformation results in \eqref{eq:transformed-C} with $j=1$ and $j=L$ terms missing. The two edge states $|g_1 \rangle, |g_{L}\rangle$ are thus left arbitrary, reflecting the dangling degrees of freedom at the edges of an SPT chain \cite{affleck1988valence,Kennedy90}. The existence of dangling states is related to an important feature of the SPT phase, viz. symmetry fractionalization at the chain edges \cite{fidkowski2011topological,Turner2011-zi}. Since operators from $j=1, L$ are missing in \eqref{eq:transformed-C}, $(X_1 , Z_1)$ and $(X_{L}, Z_{L})$ are the obvious candidates for edge operators. In the original basis they become 
\begin{align}
{\cal U}_C X_1 {\cal U}_C^\dag & = X_1 Z_2 \equiv {\cal L}_1 \nonumber \\ 
{\cal U}_C Z_1 {\cal U}_C^\dag & = Z_1 \nonumber \equiv {\cal L}_2 \\ 
{\cal U}_C X_{L} {\cal U}_C^\dag & = Z_{L-1}X_{L} \equiv {\cal R}_1 \nonumber \\ 
{\cal U}_C Z_{L} {\cal U}_C^\dag & = Z_{L} \equiv {\cal R}_2 .  \label{eq:LR-in-cluster-H} 
\end{align}
These operators satisfy the algebra
\begin{align} \label{eq:algebra}
{\cal L}_1 {\cal L}_2 = \omega^{-1} {\cal L}_2 {\cal L}_1 , ~~ {\cal R}_1 {\cal R}_2 = \omega^{-1} {\cal R}_2 {\cal R}_1 ,
\end{align}
and commute with the open-chain cluster Hamiltonian, thereby generating the $N\times N$-fold degenerate ground states. When acting on the ground states $|\Psi_C \rangle$, the symmetry operators can be decomposed into the products of these edge operators
\begin{align}
&Q_1|\Psi_C\rangle={\cal L}_1\cdot{\cal R}_2^\dagger|\Psi_C\rangle \nonumber 
\\
&Q_2|\Psi_C\rangle={\cal L}_2^\dagger\cdot{\cal R}_1|\Psi_C\rangle.
\end{align}
Because of \eqref{eq:algebra}, the global symmetries are realized projectively on the edges. 

Another characteristic of a non-trivial one-dimensional SPT phase is the degeneracy in the entanglement spectrum \cite{entanglement_spec}. The reduced density matrix of the ground state \eqref{eq:GS-cluster-H} on a finite region $A$ consisting of sites $1\leq j\leq l$ is 
\be
\rho_A =\mcu_A \left(\sum_{\mathbf{g},\mathbf{h} \in (\mathbb{Z}_N )^{\otimes \ell} } \delta_{g_1,h_1}\delta_{g_{\ell},h_{\ell}}  |\mathbf{g}\rangle \langle \mathbf{h}|\right)\mathcal{U}^\dagger_A
\ee 
where $\mathbf{h}=\{h_1,h_2,...,h_{\ell}\}$ and  ${\cal U}_A$ is a product of the pairwise unitary similar to \eqref{eq:U_M}, tailored to the region $A$. The eigenvectors of $\rho_A$ all have zero eigenvalues except for the $N^2$ eigenvectors 
\begin{align}
\k{\psi_{h_1,h_{\ell}}} = \mcu_A \left(  \sum_{h_2,...,h_{\ell-1}} |\mathbf{h}\rangle \right) 
\end{align}
which have eigenvalue $1/N^2$ and are parameterized by $h_1,h_{\ell}\in\mathbb{Z}_N$. This $N^2$ degeneracy in the entanglement spectrum is related to the $N^2$ ground state degeneracy on an open chain. 

To summarize, the ground state of the cluster Hamiltonian is an example of topological paramagnetic state, preserving the $\mathbb{Z}_N \times \mathbb{Z}_N$ symmetries of the Hamiltonian and displaying short-range entanglement as well as symmetry fractionalization at the edges. 
In the decorated domain wall picture of SPT phases~\cite{chen14}, the ground state can be visualized as being obtained from the proliferation of domain walls (created with the operators $Z^\dag_{2j-2} Z_{2j}$ and $Z_{2j-1} Z^\dag_{2j+1}$) decorated with a $\mathbb{Z}_N$ charge contributed by the $X$ operators appearing in the $b_j$.

\subsection{No-go theorem}
Thus far we have reviewed known properties of the cluster model. One can argue that one particular property is {\it unavoidable}, namely, that for $N>2$ the cluster model is not translation-invariant. This is notable since the novel dipolar $\mathbb Z_N \times \mathbb Z_N$ SPT model which we introduce in the next section is, in fact, translation-invariant.

In the remainder of this section, we make the above claim precise by proving the following theorem:
\begin{shaded}
\textbf{Theorem 1.} Let $\ket{\varphi}$ be {\it any} translation-invariant short-range entangled state which is symmetric under $Q_1$ and $Q_2$ \eqref{eq:charge-symmetry}. Then $\ket{\varphi}$ is in the following $\mathbb Z_N \times \mathbb Z_N$ SPT class w.r.t. $Q_1,Q_2$:
\begin{itemize}
    \item If $N$ is odd, then $\ket{\varphi}$ is in the trivial SPT phase.
    \item If $N$ is even, then $\ket{\varphi}$ is trivial {\it or} in the class $[N/2] \in H^2(\mathbb Z_N \times \mathbb Z_N,U(1))=\mathbb{Z}_N$.
\end{itemize}
\end{shaded}
We note that a constructive example of the non-trivial SPT phase mentioned for even $N$ is $H = - \sum_j \left( Z_{j-1}^{N/2} X_j Z_{j+1}^{N/2} + h.c. \right)$ which has a translation invariant ground state $|\Psi_C \rangle \propto \sum_{\mathbf{g} \in (\mathbb{Z}_N )^{\otimes L} } (-1)^{\sum_j g_{j}  g_{j+1}   } |\mathbf{g}\rangle$ \cite{geraedts2014exact}. This is in a phase distinct from the cluster model defined in Eq.~\eqref{eq:cluster-H}, which corresponds to $[1] \in H^2(\mathbb Z_N \times \mathbb Z_N,U(1))=\mathbb{Z}_N$ in the group cohomology classification~\cite{pollmann10,Turner2011-zi,fidkowski2011topological,Chen2011-et,Schuch2011-jx}. For instance, the former model only has a two-fold protected edge degeneracy, in contrast to the $N$-fold degenerate edge of the cluster model in Eq.~\eqref{eq:cluster-H}. 

To prove the above theorem, we use that since $\ket{\varphi}$ is a short-range entangled symmetric state, there exists \cite{sop2012} an exponentially localized operator $\mathcal O_n$ (with support\footnote{One might naively expect support also on $m>n$, but this can be avoided by using Schmidt decomposition arguments as in Ref.~\onlinecite{sop2012}.} on $m \leq n$) such that the string operator
\begin{equation}
\mathcal S_{2n} = \cdots X_{2n-5} X_{2n-3} X_{2n-1} \mathcal O_{2n}
\end{equation}
leaves the ground state invariant, i.e., $\mathcal S_{2n} \ket{\varphi} \propto \ket{\varphi}$. The SPT class $[k] \in H^2(\mathbb Z_N \times \mathbb Z_N, U(1))$ is encoded in $\prod X_n^\dagger \mathcal O_n \prod X_n = \omega^k \mathcal O_n$. The fixed-point cluster model has $\mathcal O_n = Z_n$, giving $k=1$, indeed corresponding to the root SPT phase $[1] \in H^2(\mathbb Z_N \times \mathbb Z_N, U(1))$.

By translation symmetry, we know the ground state must also be invariant under
\begin{equation}
\mathcal S_{2n+1} = \cdots X_{2n-4} X_{2n-2} X_{2n} \mathcal O_{2n+1}.
\end{equation}
By multiplying these two string operators, we obtain that the ground state is invariant under
\begin{equation}
\cdots X_{2n-4} X_{2n-3} X_{2n-2} X_{2n-1} \left( Q_2^\dagger \mathcal O_{2n} Q_2 \mathcal O_{2n+1} \right).
\end{equation}
I.e., this is a semi-infinite string of $\prod X$ with endpoint operator $Q_2^\dagger \mathcal O_{2n} Q_2 \mathcal O_{2n+1} $. The charge of this endpoint operator under $\prod_j X_j$ is clearly $\omega^{2k}$. If $2k \neq 0 \mod N$, then this would mean that we have a non-trivial SPT phase protected by the $\mathbb Z_N$ symmetry generated by $\prod_j X_j$. However, $H^2(\mathbb Z_N,U(1)) = 0$, and thus $2k = 0 \mod N$. This proves the above theorem.

\section{Dipolar SPT}
\label{sec:3} 

\ss{The dipolar SPT model}
We now introduce a model with $\mathbb{Z}_N \times \mathbb{Z}_N$ SPT order where one of the $\mathbb{Z}_N$ symmetries will be {\it dipolar} in character. The other $\zn$ symmetry will be monopolar in nature (viz. will have generators which act identically on each unit cell), and will be referred to in the following as the {\it charge} symmetry. 

Like the conventional cluster state introduced in the previous section, the dipolar SPT is a
$\mathbb{Z}_N$ spin model defined on a 1D chain. On an infinite chain, the charge ($Q$) and dipole ($D$) symmetries are defined as 
\begin{align}
Q= \prod_j X_j , ~~ D = \prod_j ( X_j )^j . \label{eq:Q-and-D}
\end{align} 
Thinking of $X_j$ as the charge operator at the site $j$, we see that $D$ measures the dipole moment of the charge distribution. 
A clear way of illustrating the modulated nature of $D$ is to look at how it behaves under translations. 
Let $\mathcal{T}$ be the operator that translates through one unit cell (which for us will be a single site). 
We then have the following algebra: 
\begin{align}
\mathcal{T}^{-1} Q \mathcal{T} = Q~,\quad \mathcal{T}^{-1} D \mathcal{T} = Q D. \label{eq:tranQD}
\end{align}
This algebra can be taken as the definition of a dipole symmetry. 

With periodic boundary conditions, the definition of dipole symmetry becomes a subtle but interesting affair that we discuss in depth in Sec.~\ref{sec:periodic}. Until then, we will only consider either infinite chains or chains with open boundary conditions. 

As we did in our review of the conventional cluster state, we will begin our discussion of SPT physics by writing down an exactly solvable model protected by $Q$ and $D$. The Hamiltonian we will focus on is
\begin{align}
H_{D} & =-\sum_{j} ( a_j + a^\dag_j ) \nonumber \\
a_j & = Z_{j-1} \big( Z^\dag_j X_j Z^\dag_j \big) Z_{j+1} ,
\label{eq:H-dipolar} 
\end{align}
where we add the parentheses to emphasize that $a_j$ is a three-body term.
As one can check, $H_D$ is a stabilizer Hamiltonian with $a_j^N=1$ and $[a_j , a_{j'} ] =0$, so that the ground state $\k{\Psi_D}$ of $H_{D}$ satisfies $a_j|\Psi_D\rangle =|\Psi_D\rangle$ for all $j$. The explicit expression for $|\Psi_D\ran$ is 
\begin{align}
 |\Psi_D \rangle  &\propto {\cal U}_D \left( \sum_{\mathbf{g}} |\mathbf{g} \rangle \right) , \nonumber \\ 
 {\cal U}_D |\mathbf{g} \rangle & = \omega^{\sum_{j} g_{j} ( g_{j+1} -g_j ) } |\mathbf{g} \rangle. 
 \label{eq:GS-dipole} 
\end{align}
Using the conditional-gate notation of Eq.~\eqref{eq:cond}, we can write ${\cal U}_D = \prod_n CZ_{n,n+1} \; \left(CZ_{n,n}\right)^\dagger$.\footnote{Note that Eq.~\eqref{eq:cond} is well-defined for $n=m$ if $\mathcal O_n$ commutes with $Z_n$. The $CZ_{n,n}$ operation acting on a single-qudit state $|g_n\rangle$ yields $\omega^{g_n^2}$, as desired.}
The model Hamiltonian $H_D$ and the ground state $|\psi_D \rangle$ are both translationally invariant and well-defined for any system size. On the other hand, there are subtleties in the {\it symmetry} of the Hamiltonian for periodic boundary conditions when the system size $L \neq 0 \mod N$, as will be discussed carefully in Sec.~\ref{sec:periodic}. 

As one can see from \eqref{eq:H-dipolar}, the stabilizers of the dipolar SPT model are constructed by decorating the charge operator $X_j$ by $Z_{j-1} Z^\dag_{j}$ on one side and $Z^\dag_j Z_{j+1}$ on the other. 
Since $Z_{j-1} Z^\dag_j$ and $Z^\dag_j Z_{j+1}$ carry opposite dipole charges under $D$, we can view $Z_{j-1} Z^\dag_{j} \cdot Z^\dag_j Z_{j+1}$ as measuring the dipole-anti-dipole charge, or as describing the domain wall of dipoles. Such a dipolar domain wall decorates the `charge' $X_j$ (this is charge with respect to $Q$, not $D$) and gives rise to the dipolar SPT model. 

For completeness of presentation, we show how to rewrite the dipolar SPT Hamiltonian in the parafermion representation in App.~\ref{app:para}.

\subsection{String order parameters and entanglement spectrum}

When placed on an infinite chain, the dipolar SPT order can be detected by the string order parameters
\begin{widetext}
\begin{align} \label{eq:string_ops} 
\mcs_Q  &= a_j a_{j+1} \cdots a_{j+m-1}  = Z_{j-1} Z^\dagger_j \( \prod_{n=1}^{m} X_{j+n-1} \) Z^\dagger_{j+m-1} Z_{j+m} \nonumber \\ 
\mcs_D & = a_j a_{j+1}^2 \cdots a_{j+m-1}^{m} = \o^{-(m^2+m)/2} Z_{j-1} \(\prod_{n=1}^{m} X^{n}_{j+n-1} \) Z^\da_{j+m-1} (Z^\da_{j+m-1} Z_{j+m})^m  . \end{align}
\end{widetext}
The string operator $\mcs_Q$ for the monopole symmetry is dressed with dipoles and anti-dipoles (created by $Z_j^\dag Z_{j+1}$ and its conjugate) at its endpoints, while the string operator $\mcs_D$ for the dipole symmetry is dressed with monopoles (viz. excitations charged under $Q$) at its edges, together with a dipole operator $(Z^\dag_{j+m-1} Z_{j+m} )^m$ whose dipole moment depends on the length of the string modulo $N$. 

Another way to detect the dipolar SPT on an infinite chain is to look for degeneracy in the entanglement spectrum. Consider a finite region $A$ consisting of sites $1\leq j\leq l$ on an infinite chain. The reduced density matrix of the ground state \eqref{eq:GS-dipole} on $A$ is
\be 
\rho_A=\mcu_A \left(\sum_{\mathbf{g},\mathbf{h}\in (\mathbb{Z}_N )^{\otimes \ell} } \delta_{g_1,h_1}\delta_{g_{\ell},h_{\ell}} |\mathbf{g} \rangle\langle \mathbf{h}|\right)\mathcal{U}^\dagger_A
\ee 
where $\mathbf{g}=\{g_1,....,g_l\}$ and $\mathcal{U}_A$ is the unitary that acts as
\begin{align}
{\cal U}_A |\mathbf{g} \rangle & = \omega^{\sum_{j=1}^{l-1}g_{j} g_{j+1}-\sum_{j=1}^{L} g_{j}^2} |\mathbf{g} \rangle  . 
\end{align}
The reduced density matrix has $N^2$ eigenvectors with a non-zero eigenvalue of $1/N^2$, given by 
\begin{align}
\k{\psi_{h_1,h_l }} ={\cal U}_A \left( \sum_{h_2,...,h_{l-1}\in\mathbb{Z}_N} |\mathbf{h} \rangle \right)
\end{align}
parametrized by $h_1,h_l\in\mathbb{Z}_N$. The entanglement entropy is then given by $2\log N$ independent of the subsystem size $l$, revealing the short-ranged entanglement of the ground state. Upon turning away from this special point the entanglement entropy will change, but the $N$-fold degeneracy will remain (as we confirm numerically in Sec.~\ref{sec:stability}). This follows from symmetry fractionalization, which we will now explain in more detail from the lens of physical edge modes. 

\subsection{Edge modes}

We now place the dipolar SPT Hamiltonian on an open chain $1\leq j\leq L$ of length $L$.
The dipolar symmetry operator $D$ is well-defined for the entire section of the open chain. Let the unitary ${\cal U}_D$ on the open chain be
\begin{align}
{\cal U}_D |\mathbf{g} \rangle & = \omega^{\sum_{j=1}^{L-1}g_{j}g_{j+1}-\sum_{j=2}^{L-1} g_{j}^2} |\mathbf{g} \rangle  . \label{eq:wf-1-open} 
\end{align}
For $j=2,... L-1$, one can easily prove the identities
\begin{align} {\cal U}_{D}^\dag X_j {\cal U}_{D} & = Z^\dag_{j-1} Z_j X_j Z_j Z^\dag_{j+1 }, \nonumber \\ 
{\cal U}_{D}^\dag a_j {\cal U}_{D} & = X_j , 
\label{eq:U_D}
\end{align} 
and use them to show
\begin{align} {\cal U}_D^\dag H_D {\cal U}_D = -\sum_{j=2}^{L-1} (X_j + X^\dag_j ) . \end{align} 

The two end sites $j=1 , L$ do not appear in the transformed Hamiltonian for an open chain,  implying that $(X_1 , Z_1)$ and $(X_L, Z_L)$ are the edge operators spanning the zero-energy subspaces. In the original basis, they become
\begin{align}
{\cal U}_{D} X_1 {\cal U}_{D}^\dag & = X_1 Z_2 ,  \nonumber \\ 
{\cal U}_{D} Z_1 {\cal U}_{D}^\dag  & = Z_1 \nonumber \\ 
{\cal U}_{D} X_L {\cal U}_{D}^\dag & = Z_{L-1} X_L  ,  \nonumber \\ 
{\cal U}_{D} Z_L {\cal U}_{D}^\dag  & = Z_L . \label{eq:dipolar-edge-1} 
\end{align} 
All the operators on the r.h.s.\ commute with $H_{D}$ and span the $N$-fold degenerate subspace at each edge. 

An alternative way to arrive at the edge operators is to concatenate all the stabilizers $a_2 \cdots a_{L-1}$ for an open chain of length $L$, which gives
\begin{align}
\prod_{j=2}^{L-1} a_j = ( Z_1 Z_2^\dag X_1^\dag  ) Q (X^\dag_L Z_L Z^\dag_{L-1} ) = 1 ,
\end{align}
the identity to 1 arising from acting on the ground state. This means the charge symmetry operator $Q$ fractionalizes to ${\cal L}_Q \cdot {\cal R}_Q$ where 
\begin{align} {\cal L}_Q =  X_1 Z_1^\dag Z_2 , ~~
{\cal R}_Q = Z_{L-1} Z_L^\dag X_L  .  \end{align}
A similar consideration for $\prod_{j=2}^{L-1} ( a_j )^j$ yields the fractionalization of dipole operator $D =\omega^{-(L+1)(L-2)/2} {\cal L}_D \cdot {\cal R}_D$ where 
\begin{align}
{\cal L}_D & = X_1 (Z_1^\dag )^2 Z_2 \sim {\cal L}_Q Z_1^\dag, \nonumber \\
{\cal R}_D & = (Z_{L-1} )^{L} (Z^\dag_L )^{L-1} ( X_L )^L \sim ( {\cal R}_Q )^L Z_L .  \label{eq:edge-algebra}    
\end{align}
Here, $\sim$ means that the l.h.s and the r.h.s are equivalent up to a phase. The edge operators we have constructed span the full space $N$-fold degenerate edge states by virtue of the algebra
\begin{align}
{\cal L}_Q {\cal L}_D & = \omega {\cal L}_D {\cal L}_Q, \nonumber \\ 
{\cal R}_Q {\cal R}_D & = \omega^{-1} {\cal R}_D {\cal R}_Q ,
\label{eq:dipolar-projective-sym}
\end{align} 
and their commutativity with the Hamiltonian $H_{D}$. 
The fractionalized charge operators ${\cal L}_Q , {\cal R}_Q$ contain the dipoles $ Z^\dag Z$, while the fractionalized dipole operators ${\cal L}_D , {\cal R}_D$ contain charges $Z$.
Such symmetry fractionalization is a robust feature, in the sense that it cannot be undone whilst preserving the finite gap of the system \cite{fidkowski2011topological,Turner2011-zi}; this projective boundary action thus protects edge modes in the entire SPT phase.

\subsection{Periodic chains and bundle symmetries}
\label{sec:periodic}

In the discussion so far we have restricted our attention to infinite or open chains. With periodic boundary conditions (PBC), the nature of the dipole symmetry becomes more subtle. On a periodic chain of length $L$, the global dipole operator $D$ is consistent with PBC $X_{j+L}=X_j$ only when $L$ is divisible by $N$, so that $X_j^{j+L}=X_j^j$. When $L$ is not a multiple of $N$, $D$ itself is no longer a symmetry; the remaining global symmetry is instead generated by $D^k$, where $k=N/{\rm gcd}(L, N)$ is the smallest integer such that $(X^k )^{L} = 1$ \cite{seiberg22a}. As a result, the global symmetry group is no longer $\mathbb{Z}_N \times \mathbb{Z}_N$, but rather $\mathbb{Z}_N \times \mathbb{Z}_{{\rm gcd}(L,N)}$. In the most extreme case of co-prime $L, N$, i.e.\ ${\rm gcd}(L, N) =1$, the dipole symmetry is absent altogether. 

\begin{figure}
\centering
\includegraphics[width=.35\textwidth]{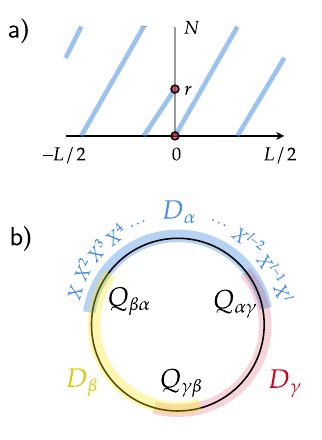} 
\caption{\label{fig:bundlesyms} Illustration of a $\zn$ dipolar bundle symmetry on a length-$L$ chain with periodic boundary conditions {\bf a)} on a chain with $r = ( L\text{ mod }N) \neq 0$, the dipole operator $D = \prod_j U_j^j$ cannot be consistently defined on the entire chain. The blue lines indicate the power of $U_j$ applied at a given position on the chain ($x$-axis). The operator is a global symmetry only when the powers at adjacent sites differ by $\pm1 \text{ mod }N$. {\bf b)} A dipole symmetry bundle is defined by splitting the spatial manifold into contractible patches $A_\a$ and defining a dipole operator $D_\a = \prod_{j \in A_\a} U^{j+n_\alpha}$ on each patch. On overlaps between patches the $D_\a$ differ by ``transition functions'' $Q_{\a\b}$ equal to powers of the charge generator $Q$}
\end{figure}

This raises the following question: can the SPT phase change---or even disappear entirely---when we work with periodic boundary conditions? On the one hand, with PBC the global symmetry group protecting the SPT phase is certainly affected by the choice of $L$. On the other hand, the ground state $\k{\Psi_D}$ and its associated string order are well-defined for all $L$, even for a periodic chain. Furthermore, it is intuitively clear that features of reduced density matrices, such as the entanglement spectrum degeneracy, should be identical with both infinite and periodic boundary conditions. This is related to the fact that while for general $L$ the dipole generator $D$ can fail to commute with $H$, it fails to commute in a very mild way, with $D$ not commuting with $H$ at only a single location in space. Moreover, the apparent position of this non-commuting location can be arbitrarily shifted 
by multiplying $D$ by integer powers of $Q$, which is itself a symmetry! Thus we may say that $D$ is a symmetry when one looks at {\it any local} patch of space, but that there is a topological obstruction when we try to {\it globally} define $D$. 

The discussion above brings to mind mathematical formulations of gauge theories, where fields are defined locally on different patches of the manifold, and transition functions on patch overlaps define how fields on different patches are related to one another. To develop vocabulary allowing us to talk about this type of situation in a more precise way, we will refer to the ``symmetry'' that arises in this context as a {\it bundle symmetry}. It is the presence of a bundle symmetry---rather than a global symmetry {\it per se}---which protects the existence of the SPT order studied in this work. We thus may refer to these phases as ``bundle symmetry protected topological phases''. 

The general construction we consider for dipolar bundle symmetries is illustrated in Fig.~\ref{fig:bundlesyms}. Consider a length-$L$ chain with PBC. To define the dipolar bundle symmetry, we divide the chain into a collection of patches $A_\a$, where we take each $A_\a$ as well as each non-empty overlap $A_\a \cap A_\b$ to be extensively large. On each patch $A_\a$ we define a dipole operator 
\be D_\a = \prod_{j \in A_\a} U^{j+n_\a}_{j},\ee 
where $j=0$ corresponds to the site on the leftmost edge of $A_\a$ and $n_\a$ is an arbitrary integer. For each pair of overlapping patches $A_\a,A_\b,$ we allow transition functions 
\be Q_{\a\b} = (D_\a^\da D_\b)|_{A_\a \cap A _\b}.\ee 
From the definition of the $D_\a$ we see that $Q_{\a\b} = (Q|_{A_\a \cap A_\b})^{n_{\a\b}}$, where the integer $n_{\a\b}$ is determined by $n_\a,n_\b$ and the distance between the endpoints of $A_\a,A_\b$. A three-patch decomposition of a chain with the associated $D_\a,Q_{\a\b}$ is illustrated in panel~b) of Fig.~\ref{fig:bundlesyms}. 

A generic local Hamiltonian is considered symmetric under the dipole bundle symmetry if the Hamiltonian supported on the patch $A_\a$ commutes with the dipole operator $D_\a$. Note that each $D_\a$ is itself not a symmetry of the full dipole-conserving Hamiltonian, since it will fail to commute with those terms in the Hamiltonian which are localized to $\p A_\a$. As the truncated Hamiltonian is symmetric with respect to each $D_\a$, it is also necessarily symmetric with the transition function $Q_{\alpha\beta}$.

To figure out if each $D_\a$ can be extended to a genuine global symmetry of the full Hamiltonian, we need to look for a dipole operator supported on a {\it single} patch that covers the full system. It is equivalent to finding a global section of the dipole bundle symmetry i.e.\ a set of patches and integers $n_\a$ which ``trivialize the bundle'', viz.\ which are such that $Q_{\a\b} = \unit$ for each pair of intersecting patches $A_\a,A_\b$. In the present example, it is easy to see that this is possible iff $L\mod N = 0$. However, such a global section is not necessary to have a notion of non-trivial SPT phases. E.g., the notion of symmetry fractionalization \cite{Turner2011-zi,fidkowski2011topological}---where applying the symmetry on a large-but-finite region leads to anomalous symmetry actions at the boundary of said region---carries over to symmetry bundles without global sections, where it is still able to protect entanglement degeneracies and string order parameters. Indeed, our dipole SPT model is such an example!

The concept of a bundle symmetry extends beyond just dipole symmetry, and in App.~\ref{app:symm_bundles} we give a more detailed treatment of how bundle symmetries can be defined. This framework includes a broader class of modulated bundle symmetries such as the quadrupolar and exponentially modulated examples studied below. 
 In the future it could be interesting to explore the topology of bundle symmetries in more detail, by e.g.\ constructing examples in higher dimensions with magnetic monopole-like topologies. Some fundamental questions worth investigating are whether such bundle symmetries can be spontaneously broken, or gauged, or even arise as an emergent low-energy property---all of which are familiar and important for usual symmetries.

\ss{Relation to cluster SPT model}

So far we have seen two models both protected by $\mathbb Z_N \times \mathbb Z_N$ symmetry. The first is the conventional cluster SPT model reviewed in Sec.~\ref{sec:2}, and the second is the dipolar SPT model introduced above. They are protected by two charge symmetries, \eqref{eq:charge-symmetry}, or by one charge and one dipole symmetries, \eqref{eq:Q-and-D}. 

For $N=2$, these models are the same up to an overall sign, since when $N=2$ we have $a_j = Z_{j-1} Z_j X_j Z_j Z_{j+1} = -b_j$. 
A natural question is what, if any, is the relation between these two types of SPT stabilizer models when $N>2$?  We have two key results to answer this question: a {\it no-go} result (theorem 2) and a constructive one (theorem 3).

\begin{figure*}
    \centering
\begin{tikzpicture}[scale=0.75]
\fill[black,opacity=0.07] (7.5,0.6) rectangle (15.5,-7.6);
\foreach \shift in {0,8,16}{
    \foreach \x in {0,1,2,3,4,5,6,7}{
        \draw[-] (\shift+\x,-7.8) -- (\shift+\x,0.8);
    };
    
    \foreach \x in {0,2,4,6}{
        \draw[red,-] (\shift+\x,0) -- (1+\shift+\x,0);
        \filldraw[red] (\shift+\x,0) circle (2pt);
        \draw[fill=white,draw=red,text=red] (1+\shift+\x,0) circle (10pt) node {$Z^\dagger$};
        
        \ifthenelse{\x<6}{\draw[-,red] (1+\shift+\x,-1) -- (2+\shift+\x,-1);
        \filldraw[red] (1+\shift+\x,-1) circle (2pt);
        \draw[fill=white,draw=red,text=red] (2+\shift+\x,-1) circle (10pt) node {$Z$};}
        
        \draw[-,blue] (\shift+\x,-2) -- (1+\shift+\x,-2);
        \filldraw[blue] (\shift+\x,-2) circle (2pt);
        \draw[fill=white,draw=blue,text=blue] (1+\shift+\x,-2) circle (10pt) node {$X$};
        
        \ifthenelse{\x>0}{\draw[-,blue] (\shift+\x,-3) -- (1+\shift+\x,-3);
        \filldraw[blue] (1+\shift+\x,-3) circle (2pt);
        \draw[fill=white,draw=blue,text=blue] (\shift+\x,-3) circle (10pt) node {$X^\x$};}
        
        \draw[-,blue] (\shift+\x,-4) -- (1+\shift+\x,-4);
        \filldraw[blue] (\shift+\x,-4) circle (2pt);
        \draw[fill=white,draw=blue,text=blue] (1+\shift+\x,-4) circle (10pt) node {$X$};
        
        \draw[-,red] (\shift+\x,-5) -- (1+\shift+\x,-5);
        \filldraw[red] (\shift+\x,-5) circle (2pt);
        \draw[fill=white,draw=red,text=red](1+\shift+\x,-5) circle (10pt) node {$Z$};
        
        \ifthenelse{\x<6}{\draw[-,red] (1+\shift+\x,-6) -- (2+\shift+\x,-6);
        \filldraw[red] (1+\shift+\x,-6) circle (2pt);
        \draw[fill=white,draw=red,text=red](2+\shift+\x,-6) circle (10pt) node {$Z$};}
        
        \draw[fill=white,draw=red,text=red] (\shift+\x,-7) circle (10pt) node[xshift=-0.5,yshift=0.5] {$R$};
        
        \draw[fill=white,draw=red,text=red] (1+\shift+\x,-7) circle (10pt) node[xshift=-0.5,yshift=0.5] {$R$};
    };
};    
\end{tikzpicture}
\caption{\textbf{Transformation between dipolar and cluster models.} As proven in the main text, there {\it cannot} exist any unitary transformation which maps the cluster model $H_C$ and its protecting symmetries $Q_1,Q_2$ to the dipolar SPT model $H_D$ and its symmetries $Q,D$. However, there does exist a unitary $\mathcal V$ which maps the {\it ground states} and protecting symmetries to one another. Moreover, this unitary is a tensor product between unit cells of size $\textrm{lcm}(N,2)$. We show it for the case $N=8$, where the gray region highlights the unit cell. The red gates are $CZ_{m,n}$ and  blue gates are $CX_{m,n}$; finally $R_n \ket{g_n} := CZ_{n,n}^\dagger \ket{g_n} = \omega^{-g_n g_n} \ket{g_n}$ is a well-defined single-qudit gate.\label{fig:unitary}}
\end{figure*}
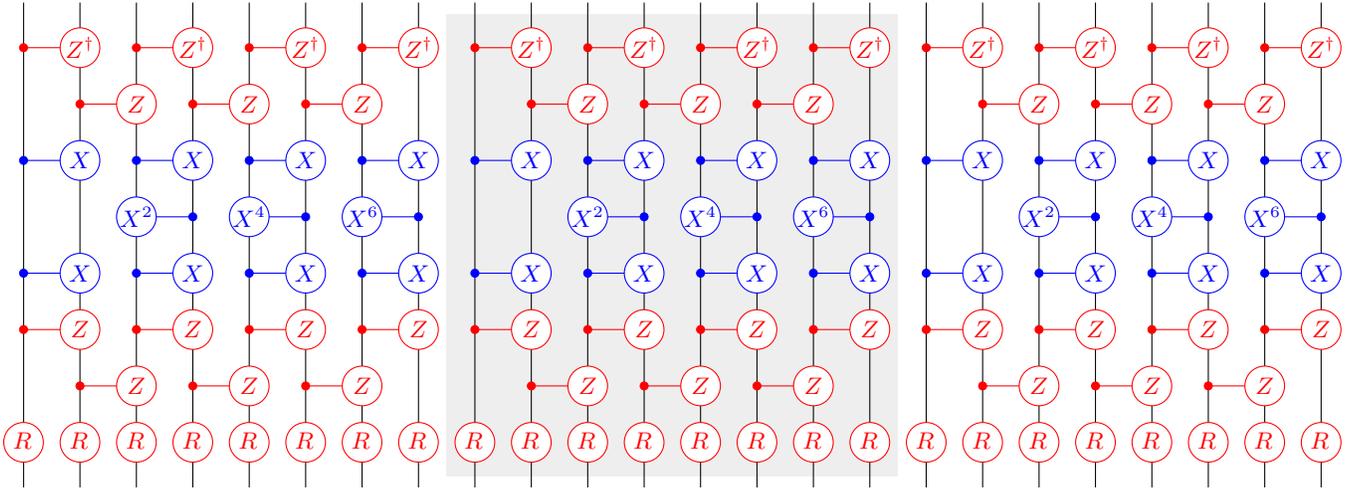

\begin{shaded}
\textbf{Theorem 2.} If $N>2$, there does {\it not} exist a unitary transformation which maps $H_D$ and its protecting symmetry group (generated by $Q$ and $D$) to $H_C$ and its protecting symmetry group (generated by $Q_1$ and $Q_2$).
\end{shaded}

This result has several practical implications. For example, it means that a perturbed Hamiltonian $H_D + \lambda H'$ respecting $Q$ and $D$ can 
not be mapped to a perturbed cluster model $H_C + \lambda H''$ respecting $Q_1$ and $Q_2$. This implies that, say, studying the criticality of phase transitions of such dipolar SPT phases cannot be reduced to the study of critical cluster chains.

The above theorem is a simple consequence of the algebraic relations between the symmetry generators and the Hamiltonian terms. In particular, note that $Q= \prod_j a_j$ and $D \propto \prod_j (a_j)^j$ (on an infinite chain for simplicity), where $a_j$ is the Hamiltonian term defined in Eq.~\eqref{eq:H-dipolar}.
In contrast, the symmetries of the cluster model are $Q_n = \prod_j b_{2j+n}$ (for $n=1,2$), where $b_j$ is defined in Eq.~\eqref{eq:cluster-H}. Hence, any unitary transformation which maps $a_j$ to $b_{k_j}$ (or its hermitian conjugate) \textit{cannot} at the same time map, say, $D\propto\prod_j a_j^j$ to $Q_1=\prod_j b_{2j-1}$ (if $N>2$).

Note that the above result does not preclude the existence of unitary transformations which map $H_D$ to $H_C$. Indeed, recall that we already wrote down finite-depth unitary circuits $\mcu_D, \mcu_C$ which map $H_C$ and $H_D$ to the same trivial Hamiltonian $-\sum_j ( X_j + X^\dag_j )$ in \eqref{eq:transformed-C} and \eqref{eq:U_D}. Hence by concatenation, 
\begin{align} {\cal U }= {\cal U}_D^{\vphantom \dagger} {\cal U}_C^\dag \end{align} 
we have ${\cal U}^\dagger a_j {\cal U} = b_j$ and ${\cal U}^\dagger H_D {\cal U} = H_C$ mapping between $H_C$ and $H_D$. Moreover, unlike the individual unitaries ${\cal U}_C$ and ${\cal U}_D$, this composite unitary is a tensor product between two-site unit cells, which implies that properties such as entanglement spectrum degeneracy are invariant under ${\cal U}$. Indeed,\footnote{If one desires a unitary which is a tensor product between unit cells $\{ (2n-1,2n)\}$ one can instead use ${\cal U}_D^{\vphantom \dagger} {\cal U}_C T$ where $T$ is complex conjugation in the $\{|g\rangle\}$ basis.}
\begin{equation}
{\cal U} = \prod_n \left( CZ_{2n,2n+1} \right)^2 \times \prod_m\left(CZ_{m,m}\right)^\dagger.
\end{equation}

However, in line with the above theorem, this unitary does {\it not} map the protecting symmetries $(Q,D)$ into $(Q_1,Q_2)$. In particular, this implies that ${\cal U}$ cannot be used to map the study of symmetric deformations of these models. For instance, it does not map the trivial dipolar SPT model $H_0 = - \sum_j ( X_j+ X_j^\dag )$ 
to the trivial monopolar SPT model (which would have the same Hamiltonian) but instead to 
\begin{align}
{\cal U}^\dagger H_0 {\cal U} &= - \sum_j \big[ Z_{2j} X_{2j} Z_{2j} \big( Z_{2j+1}^\dagger\big)^2 \nonumber \\
& \quad \qquad + \big( Z_{2j}^\dagger\big)^2 Z_{2j+1} X_{2j+1} Z_{2j+1} + h.c. \big]
\end{align}
While this model respects total charge symmetry, it indeed does not commute with $Q_1$ or $Q_2$. Hence, studying the interpolation from our dipolar SPT model $H_D$ to the trivial Hamiltonian $H_0$ (which we study in Sec.~\ref{sec:stability}) does {\it not} correspond to some previously-studied perturbation of the cluster model (for $N>2$).

Since the above unitary transformation from the dipole SPT chain $H_D$ to the monopole SPT chain $H_C$ does not map the dipole symmetry $D$ to one of the sublattice charge symmetries, it raises the question: What does it map to? We find the answer is
\begin{equation}
{\cal U}^\dagger D \; {\cal U} \; \propto \; D \times \prod_j \big( Z_{2j-1}^{\vphantom \dagger} Z_{2j}^\dagger \big)^2 \propto \prod_j (b_j)^j. \label{eq:newD}
\end{equation}
Indeed, this commutes with the cluster SPT model $H_C$, and by virtue of the above mapping we discover that $H_C$ is not just a non-trival SPT model for the usual $Q_1,Q_2$ symmetry but {\it also} for total charge $Q$ and the `dipolar' symmetry shown on the r.h.s. of \eqref{eq:newD}.

Our second key result in this subsection shows that by forgetting about energetics and focusing on the ground states, there does exist a mapping, which is albeit rather subtle: 
\begin{shaded}
\textbf{Theorem 3.} There exists a unitary transformation which maps:
\begin{itemize}
\item[(i)] {\it the ground state} of $H_D$ to {\it the ground state} of $H_C$
\item[(ii)] the symmetry generators $Q$ and $D$ to $Q_1 Q_2^\dagger$ and $Q_1$, respectively.
\end{itemize}

Moreover, such a unitary can be chosen to be a tensor product, but for $N>2$ it must have a unit cell of size at least $l \equiv \textrm{lcm}(2,N)$.
\end{shaded}

The fact that such a unitary can be chosen to be a tensor product unitary implies it does not change the SPT properties (e.g., entanglement degeneracies) between these unit cells. However, the fact that the unit cell grows linearly with $N$ illustrates that identifying the wavefunctions requires a considerable restructuring of the degrees of freedom. This minimal size of the unit cell follows from the fact that the symmetries satisfy different algebras with the translation operator unless the translation is by a multiple of $l$.

We now construct the unitary claimed in theorem 3. For notational convenience, we just define the unitary for the unit cell of $l$ qudits; the global unitary is then a tensor product over unit cells. This unitary consists of three layers $\mathcal V = {\mathcal V}_3 {\mathcal V}_2 {\mathcal V}_1 $, which are defined as follows:\begin{align}
{\mathcal V}_1 &= \prod_{n=1}^{l/2} CZ_{2n-1,2n}^\dagger \prod_{n=1}^{l/2-1} CZ_{2n,2n+1} \nonumber \\
{\mathcal V}_2 &= \prod_{n=1}^{l/2} CX_{2n-1,2n} \left(CX_{2n,2n-1}\right)^{2n-2} CX_{2n-1,2n}  \nonumber \\
{\mathcal V}_3 &= \prod_{n=1}^{l-1} CZ_{n,n+1} \prod_{n=1}^l CZ_{n,n}^\dagger.
\end{align}
We visually represent the circuit in Fig.~\ref{fig:unitary} for the case $N=8$. A direct computation shows that it maps the stabilizers as follows:
\begin{align}
\mathcal V b_{2j-1} {\mathcal V}^\dagger &= \left(a_{2j-1}\right)^{2j-1} \left(a_{2j}\right)^{2j} \nonumber \\
\mathcal V b_{2j} {\mathcal V}^\dagger &= \left(a_{2j-1}\right)^{2j-2} \left(a_{2j}\right)^{2j-1} .
\end{align}
From this, we directly see it maps the ground states to one another\footnote{This follows from: (i) $\mathcal V$ clearly maps the ground state subspace of $H_D$ to that of $H_C$, (ii) both g.s.\ spaces are isomorphic (e.g., unique g.s.\ on periodic boundary conditions), and (iii) $\mathcal V$ is a unitary map.}, since these are characterized by $a_j\ket{\psi_D} = \ket{\psi_D}$ and $b_j \ket{\psi_C} = \ket{\psi_C}$. Moreover, it maps the symmetries as follows:
\begin{align}
\mathcal V Q_1 \mathcal V^\dagger &= \prod_j \mathcal V b_{2j-1} {\mathcal V}^\dagger = \prod_j a_j^j = D \nonumber \\
\mathcal V Q_2 \mathcal V^\dagger &= \prod_j \mathcal V b_{2j} {\mathcal V}^\dagger = \prod_j a_j^\dagger \prod_j a_j^j = Q^\dagger D .
\end{align}
We have thus confirmed the properties claimed in theorem 3. We stress again that this unitary does {\it not} map the respective Hamiltonians. Indeed,
\begin{equation}
\mathcal V H_C {\mathcal V}^\dagger = - \sum_j \sum_{p=0,1} \left( a_{2j-1}^{2j-1-p} a_{2j}^{2j-p} + h.c. \right).
\end{equation}
Although this fails to be translation-invariant and has quite different energetics to $H_D$, we directly see it does share the same ground state as $H_D$.

\ss{Other dipolar SPT phases}

So far we have discussed the SPT phase in which the charge operator $X$ is decorated with operators that carry unit dipole moment. The remaining dipolar SPT phases are obtained by decorating the charge operator with operators possessing dipole moments of $\eta\in \zn$. The stabilizer Hamiltonian for a given value of $\eta$ is obtained through an appropriate modification of the stabilizers $a_j$ as  
\begin{align} H_\eta & = - \sum_j ( a_j^{(\eta)}  + [ a_j^{(\eta)} ]^\dag )  , \nonumber \\ 
a_j^{(\eta)} & = ( Z_{j-1} Z^\dag_j )^\eta X_j ( Z^\dag_j  Z_{j+1} )^\eta .  \label{eq:H_eta}
\end{align} 
Its ground state is 
\begin{align} \k{\Psi_\eta} & = {\cal U}_D^{(\eta)}  \left( \sum_{\bfg} |\bfg \ran \right) \nonumber \\ 
{\cal U}_D^{(\eta)}  |\bfg \ran  & = \o^{\eta \sum_j (g_j g_{j+1} -g_j^2) } |\bfg\ran. \label{eq:psi-eta} \end{align}
To prove this, it suffices to show $(\mcu_D^{(\eta)})^\dag a_j^{(\eta)} \mcu_D^{(\eta)} = X_j$. The global symmetries of $H_\eta$ are still given by $Q$ and $D$ of \eqref{eq:Q-and-D}.

The fractionalized symmetry operators at the edges are now given by 
\bea \label{general_edge_alg}
{\cal L}_Q^{(\eta)} &= X_1 ( Z_1^\dag  Z_2 )^\eta , \\ 
{\cal R}_Q^{(\eta)} &= (Z_{L-1} Z_L^\dag )^{\eta} X_L  ,  \\
{\cal L}_D^{(\eta)} &= X_1 (Z_1^\dag )^{2\eta} Z_2^\eta  , \\  
{\cal R}_D^{(\eta)} &=  (Z_{L-1} )^{\eta L} (Z^\dag_L )^{\eta (L-1)} ( X_L )^L . \eea 
Note ${\cal L}_D^{(\eta)} \sim {\cal L}_Q^{(\eta)} (Z_1^\dag )^\eta$ and ${\cal R}_D^{(\eta)} \sim  ( {\cal R}_Q^{(\eta)} )^L (Z_L)^\eta$. The edge algebra then becomes
\begin{align}
{\cal L}_Q^{(\eta)} {\cal L}_D^{(\eta)} & = \omega^\eta {\cal L}_D^{(\eta)} {\cal L}_Q^{(\eta)} , \nonumber \\ 
{\cal R}_Q^{(\eta)} {\cal R}_D^{(\eta)} & = \omega^{-\eta } {\cal R}_D^{(\eta)} {\cal R}_Q^{(\eta)} . 
\label{eq:eta-algebra} 
\end{align}
This edge algebra enforces the minimal number of protected edge modes to be 
\begin{align} d_\eta \equiv N/\gcd(N,\eta). \label{eq:d-eta} \end{align} 

The Hamiltonian written down in~\eqref{eq:H_eta} however has $N$ degenerate states per edge, as it transforms to $-\sum_{j=2}^{L-1} (X_j + X_j^\dag )$ under ${\cal U}_D^{(\eta)}$. The degeneracy can be lifted to be the minimal allowed value $d_\eta$ by adding the following edge terms 
\begin{align}
   \Delta H=-\left[(\mathcal{L}_{Q}^{(\eta)})^{d_\eta}+(\mathcal{R}_{Q}^{(\eta)})^{d_\eta}+h.c.\right]
   \label{eq:edge_hamiltonian_dipole}
\end{align}
which commute with the stablizers $a_j^{(\eta)}$ and preserve the dipole symmetry (note that $(\mcl^{(\eta)}_Q)^{d_\eta} = X_1^{d_\eta}$ and $(\mcr^{(\eta)}_Q)^{d_\eta} = X_L^{d_\eta}$). A simple way to see how the degeneracy lifting occurs is to work in the ground state basis $|n_L,n_R\rangle$, $n_{L,R}\in\mathbb{Z}_N$ where $\mathcal{L}_{Q}^{(\eta)}$ and $\mathcal{R}_{Q}^{(\eta)}$ are diagonalized with quantum numbers $\omega^{n_L}$ and $\omega^{n_R}$, respectively. With the $\Delta H$, the energy levels will split to $ -2 \cos [ 2\pi n_L d_\eta / N]-2 \cos [ 2\pi n_R d_\eta / N]$ and the new ground states have $n_{L,R}$ divisible by $N/d_\eta$. This leaves behind a $d_\eta$-fold degeneracy per edge. 

	\subsection{MPS representation \label{subsec:MPS}}
	
	The ground states of 1D SPT phases---being weakly entangled---admit compact matrix product state (MPS) representations \cite{Fannes1992,bridgeman17,Cirac21}. In the study of conventional 1D SPT phases, the MPS formalism has the merit of making the symmetry fractionalization pattern and nature of the degenerate edge modes quite explicit, and was one of the approaches used to originally define and classify such phases \cite{Chen2011-et,Pollmann2012-lv}. In this section we will see how MPS techniques can similarly be used to understand and classify dipolar SPTs. 

 \sss{General dipole symmetries}
 
	Before specializing to the context of the $\zn$ dipole symmetry as studied above, we first consider a general internal charge symmetry group $G$ and its associated dipole symmetry group $G_D$. For a given $g\in G$, we let the associated actions by $G$ and $G_D$ be represented in terms of tensor products of single-site unitaries as 
 \begin{align} U_Q(g) = \prod_{j=1}^L U_j (g), ~~
 U_D(g) = \prod_{j=1}^L( U_j (g) )^j, \label{eq:charge-and-dipole-unitary} \end{align}
 respectively. Note that the same on-site unitary $U_j (g)$ appears in both of the global symmetry operators.  
	
Consider a weakly entangled translation-invariant wave function $\k\Psi$, with $\k\Psi$ transforming trivially under $G$ and $G_D$: $U_Q (g) |\Psi\rangle = e^{i\gamma_Q(g)} |\Psi\rangle$, $U_D (g) |\Psi\rangle = e^{i\gamma_D(g)}|\Psi\rangle$ for some phases $\gamma_Q(g), \gamma_D(g)$. The state $\k\Psi$, being weakly entangled and translation-invariant, can be expressed in MPS form as 
\be \label{mpsrep} \k{\Psi} = \sum_{\bfa \in \zz^L_N} \Tr[BA^{a_1} \cdots A^{a_L} ]\k{\bfa}, \ee
	where we have assumed an $N$-dimensional on-site Hilbert space labeled by $a \in \zn$ instead of $g$ as in previous sections to avoid notational overlap with the group element $g$. The matrix $B$ fixes the boundary conditions of the MPS, with $B = \unit$ for the periodic boundary conditions  and $B = \k{\psi_R}\bra{\psi_L}$ for an open-chain MPS whose virtual indices are fixed as $\k{\psi_{L/R}}$ on the left/right ends. Due to the bulk $G$-invariance of $\k \Psi$, the fundamental theorem of MPS \cite{bridgeman17,Cirac21} mandates that the tensors $A^a$ obey 
\be \label{fundtheorem} \sum_{b\in \zn} [U(g)]^{ab} A^b = e^{i\theta_g } V(g)^\dagger A^a V(g)\ee 

	for some set of unitary matrices $V(g)$ and phases $e^{i\theta_g}$. For notational convenience we will omit $e^{i\theta_g}$ below. Graphically, \eqref{fundtheorem}
 reads 
\be \igptc{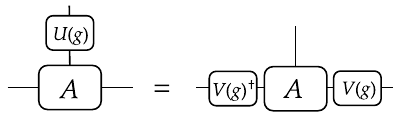} \ee 
 
 The condition \eqref{fundtheorem} then dictates that the dipolar action on $\k\Psi$ be 
	\be 
 \label{udonpsi} 
 U_D(g) \k\Psi = \sum_{\bfa \in \zz^L_N} \Tr[ B  V(g)^\da A^{a_1} \cdots V(g)^\dagger A^{a_L} V(g)^L ]\k{\bfg}, \ee 
in effect replacing the action of $U_D (g)$ on physical indices by the transformation of MPS tensors $A^a \rightarrow V(g)^\da A^a$. 
In particular, the virtual action of $G_D$ is entirely fixed by that of $G$ itself. It is then interesting to understand how a mixed anomaly between $G$ and $G_D$ can arise even in the absence of a self-anomaly for $G$, as is the case in the $G=\zn$ examples studied above.

From~\eqref{udonpsi}, we see that the requirement that $\k\Psi$ be bulk $G_D$-invariant imposes the nontrivial restriction that \eqref{mpsrep} be invariant under insertion of $V(g)^\da$ on every virtual bond of the MPS. In turn, this can be rephrased as the requirement that for all $g\in G$,  
	\be V(g)^\da A^a  = e^{i \phi_g} V'(g)^\da A^g V'(g)  \label{eq:V-and-V'} \ee 
holds for some set of unitaries $V'(g)$ and phases $e^{i\phi_g}$. Again omitting the phases, this equation can be written graphically as 
\be \igptc{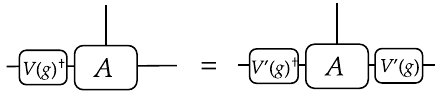} \ee 

While the relation \eqref{fundtheorem} applies to all MPS tensors representing a translation-symmetric state, the other relation \eqref{eq:V-and-V'} arises from both the charge and the dipole symmetries being represented by utilizing the same on-site unitary $U_j (g)$ as in \eqref{eq:charge-and-dipole-unitary}. 

In conventional SPT phases, it is the $V(g)$'s that form a projective representation of $G$ with $V(g) V(h) = \o(g,h) V(gh)$ for some cohomology class $\o \in H^2(G,U(1))$, and lead to fractionalization. 
In the dipole SPT phases, for a general finite abelian group $G$, $G_D$, the projectivity arises from the nontrivial commutation relations between the $V(g)$ and the $V'(g)$:
\begin{align} V(g)V'(g)V(g)^\da V'(g)^\da \neq 1 . \end{align} 

With $V(g)$ and $V'(g)$ in place for a given $A^a$, the action of the symmetries $U_Q(g), U_D(g)$ on an open chain of length $L$ with $B = | \psi_R \rangle \langle \psi_L |$ results in $\k{\psi_{L/R}}$ transforming according to the operators 
	\begin{alignat}{2} \mcl_Q(g) & =V(g), \quad&&\mcr_Q(g)  = V(g),\\ 
	\mcl_D(g) & = V'(g),\qquad
	&&\mcr_D(g)  = V'(g)V^L(g) , \nonumber
	\end{alignat}
e.g. $|\psi_L \rangle \rightarrow V(g) |\psi_L \rangle$ under $U_Q (g)$.

Let us define the operators $U_L (g), U_R (g)$ by 
\begin{align}
&\sum_b [U_L (g)]^{ab} A^b= V(g)^\da A^a,\nonumber \\
&\sum_b [U_R (g)]^{ab} A^b = A^a V(g),
\end{align}
so that the on-site symmetry action factorizes as $U_Q(g) = U_L (g) U_R (g)$. Similarly, let us also define $U_L' (g), U_R' (g)$ by
\begin{align}
&\sum_b [U_L' (g)]^{ab} A^b= V'(g)^\da A^a,\nonumber \\
&\sum_b [U_R' (g)]^{ab} A^b = A^a V'(g) . 
\end{align}
Using them, the string operators for the $G$ and $G_D$ symmetries can be constructed by decorating symmetry operators on a finite interval by appropriate factors of $U_{L/R}(g)$ at the edges as 
\bea &\mcs_Q(g) = U_R(g)_j \( \prod_{n=1}^{m-1} U_Q(g)_{j+n} \) U_L(g)_{j+m} \nonumber \\ 
&\mcs_D(g)  =  
\\
&{U_R'}(g)_j \( \prod_{n=1}^{m-1} ( U_Q(g)_{j+n} )^{j+n} \) {U_L'}(g)_{j+m}( U_L(g)_{j+m} )^{j+m} . 
\eea 
The $U_L , U_R$ and $U'_L , U'_R$ operators are placed in such a way that the actions of both string operators on the MPS wave function \eqref{mpsrep} is an identity operation. 

\sss{$\zn$ dipole symmetry}

Having discussed the general structure of the virtual symmetry action in a dipolar SPT, we now return to the specific case of $G = \zn$ that has been the focus of our attention above. The notation $\k\bfg$ is accordingly restored to label Hilbert space basis vectors. To rewrite the dipolar SPT state \eqref{eq:GS-dipole} in an MPS representation, it is helpful to define the following matrices: 
\bea  R & \equiv \sum_{g\in \zn} \o^{-g^2} \proj{g}  \\ 
		W_\eta & \equiv \sum_{g,h \in \zz_N} \o^{\eta gh} |g\rangle \langle h | ,
	\eea 
where again $\eta \in \mathbb{Z}_N$ labels different dipolar SPT phases. One can show that they satisfy the algebra
	\bea  
 \label{wandt}
		XR & = R Z XZ,\quad ZR = RZ ,  \\ 
		 W_\eta Z^\eta & = X^\dag W_\eta, ~~ \quad W_\eta X = Z^\eta W_\eta . 
	\eea 
 One recognizes $W_\eta$ as the matrix that implements the $\eta$-twisted discrete Fourier transform. Note that $W_\eta$ is unitary only when $d_\eta = N$ (i.e. ${\rm gcd}(N, \eta ) = 1$, see \eqref{eq:d-eta}), in which case it squares to charge conjugation: $W_\eta^2 = C \equiv \sum_g |g \rangle \langle -g  |$. 

	MPS tensors for the ground state wave function \eqref{eq:psi-eta} can then be written down as 
	\be A_\eta^g = \o^{-\eta g^2}W_\eta \proj{g}.\ee 
	In terms of tensor diagrams, we have 
\be \igptc{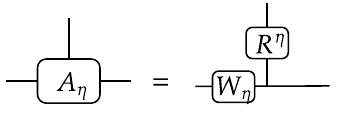} .\ee 
Using the identities in \eqref{wandt} as well as 
 \be \igptc{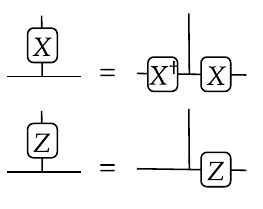},\ee
One can check that the MPS tensor satisfies the identity 
\begin{align} [X]^{gh} A_\eta^h = (XZ^\eta)^\da A_\eta^g (XZ^\eta) . \end{align} 
Comparing it to the fundamental MPS theorem of \eqref{fundtheorem}, the fractionalization of the charge symmetry is implemented by 
\be V_\eta\equiv XZ^\eta\ee 
in the dipolar SPT model. 

We now determine the matrix $V'_\eta$ which fixes how the {\it dipole} symmetry fractionalizes. This can be done by using the identity 
\be X^\da A_\eta^g Z^{-\eta} = A_\eta^g,\ee 
which is easily proved using \eqref{wandt}. This identity can be rewritten as 
\be (XZ^\eta)^\da A_\eta^g = (Z^\eta)^\da A_\eta^g Z^\eta,\ee  
 which tells us that $V' = Z^\eta$. The commutation relations between 
 \begin{align} V = XZ^\eta , ~~ V' = Z^\eta \nonumber \end{align} then reproduce the edge mode algebra derived earlier in \eqref{eq:eta-algebra}. 

 We now discuss the edge degeneracy when the system is placed on an open chain so that $B = |{\psi_R}\rangle \langle{\psi_L}|$. When $d_\eta = N$, the action generated by $V,V'$ forms an irreducible projective representation of $\zn\times \zn$. Such irreps always have dimension $N$, and the edge mode degeneracy cannot be reduced below $N$. When $d_\eta < N$ however, the story is different.\footnote{Note that in this case, the non-unitarity of $W_\eta$ means that the MPS is not injective after blocking two sites. More detailed discussions of how the representation theory and MPS tensors work out, in this case, can be found in \cite{stephen2017computational} and App. H of \cite{lake2022exact}. For the present purposes, a full discussion of these issues would take us too far afield.} 
In this case, both $V, V'$ commute with $X^{d_\eta}$. This means that without violating either charge or dipole symmetries, we can insert projectors onto the $+1$ eigenspace of $X^{d_\eta}$ on both the leftmost and rightmost virtual legs of the MPS (which fix the boundary conditions). More explicitly, we may replace the matrix $B$ which fixes the boundary conditions by $\Pi_\eta B \Pi_\eta$, where 
\be \Pi_\eta = \frac1{\gcd(N,\eta)}\sum_{n = 1}^{\gcd(N,\eta)} X^{n d_\eta}.\ee 
Since $\dim\, {\rm Im}\, \Pi_\eta = d_\eta$, we recover the edge degeneracy of $d_\eta$ as argued for above on the grounds of the relations obeyed by the string operators. 

\begin{figure*}
    \centering
    \begin{tikzpicture}
    \node at (0,0){
        \includegraphics[scale=0.32]{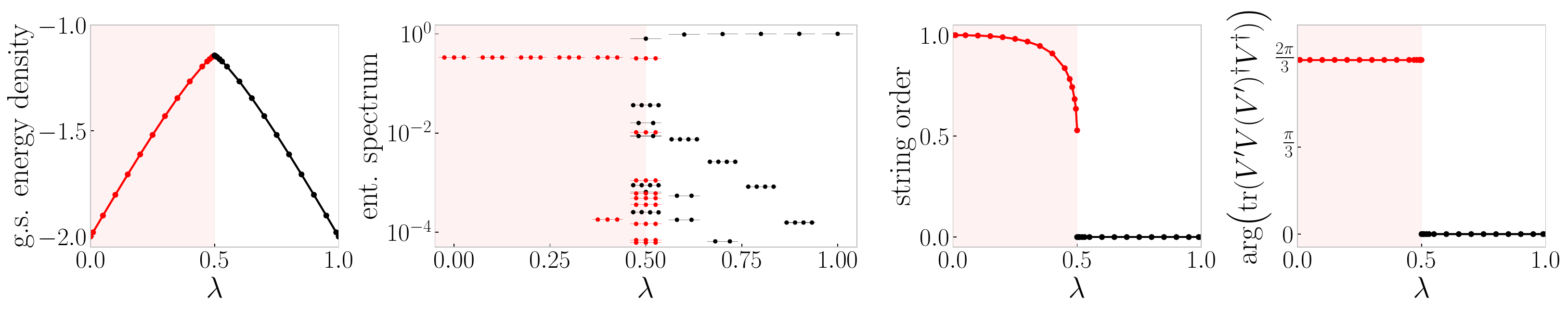}
    };
    \node at (-7.4,0.8){(a)};
    \node at (-3.5,0.8){(b)};
    \node at (2.3,0.8){(c)};
    \node at (6.3,0.8){(d)};
    \end{tikzpicture}
    \caption{\textbf{Stability and phase transition of $\mathbb Z_3$ dipolar SPT model.} We interpolate from the dipolar SPT model to the trivial Hamiltonian. (a) The ground state energy density signals a first-order transition at $\lambda=0.5$. (b) We see that in the entire SPT phase, the entanglement spectrum is (at least) threefold degenerate, whereas in the trivial phase certain entanglement levels are unique. At the first order point $\lambda =0.5$, we show the results obtained for the twofold degenerate ground state: the SPT state (red) and the trivial state (black). (c) The topological string order parameter is nonzero in the SPT phase, and discontinuously jumps to zero when tuning across the first order transition into the trivial phase. (d) The numerically obtained quantized SPT invariant developed in Sec.~\ref{subsec:MPS} confirms the predicted property $V' V(V')^\dagger V^\dagger =\omega \cdot \unit$ in the SPT phase and $V' V(V')^\dagger V^\dagger =\unit$ in the trivial phase; in the numerics we compute $V,V'$ explicitly by diagonalizing the transfer matrix.}
    \label{fig:Z3stability}
\end{figure*}
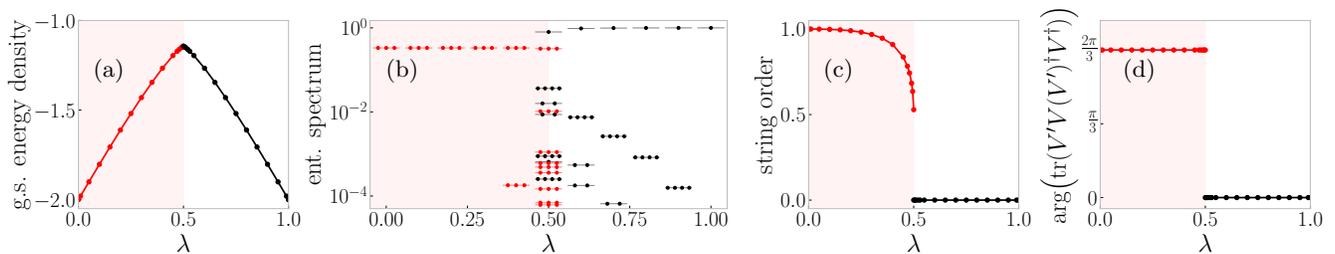

\subsection{Stability analysis \label{sec:stability}}

Thus far, we have analytically shown that the dipolar SPT model $H_D$ has symmetry-protected features such as entanglement degeneracies and string order. Here we briefly demonstrate and confirm this stability via a numerical analysis of a perturbed Hamiltonian and its quantum phase transition. 

As a minimal example, we consider the addition of the Zeeman term:
\be H = (1-\lambda) H_D - \lambda \sum_i (X_i +X_i^\dagger), \label{eq:perturbed} \ee
and study the evolution of the ground state while increasing $\lambda$. Note that $\lambda=0$ is the solvable SPT point, and $\lambda=1$ is a trivial product state. This model exhibits a $\mathbb Z_2^T$ duality: by conjugating the Hamiltonian with the unitary map ${\cal U}_D$ developed in 
\eqref{eq:GS-dipole} and subsequently performing complex-conjugation in the $|g\rangle$ basis, we effectively map $\lambda \to 1-\lambda$. Hence, if there is a direct transition between the SPT and trivial phase, it must occur at the self-dual point $\lambda=1/2$. Alternatively, there can be an intermediate phase.

We analyze the ground states of Eq.~\eqref{eq:perturbed} with $N=3$ using the infinite density matrix renormalization group (iDMRG) \cite{White92,White93} method in the open-source TeNPy python library \cite{tenpy}. This allows us to directly obtain the ground state wavefunction in the limit of an infinitely long chain, described by a translation-invariant matrix product state. Operationally, we fix a bond dimension $\chi$ associated with this MPS, and we find that for $\chi=100$ all physical quantities we consider converge. The results are shown in Fig.~\ref{fig:Z3stability}.

The ground state energy density in Fig.~\ref{fig:Z3stability}(a) is the first indication that there is indeed a direct transition at $\lambda=1/2$, and the kink suggests it is first-order. This is confirmed in the entanglement spectrum (Fig.~\ref{fig:Z3stability}(b)) where we see the robust threefold degeneracy for each level as predicted by the SPT phase, up to $\lambda=1/2$. At this self-dual point, iDMRG finds {\it two} ground states, consistent with a first-order transition. This second state connects to the region $\lambda \geq 1/2$, where its entanglement smoothly transitions into the product state at $\lambda=1$. This first-order transition between the SPT and trivial phase is also consistent with Fig.~\ref{fig:Z3stability}(c), where we find a discontinuous jump of the string order parameter at $\lambda=1/2$. Finally, since our state is translation-invariant and numerically described by a matrix product state, we can also calculate the {\it quantized} invariant introduced in Sec.~\ref{subsec:MPS}. This quantization condition is very well captured by the numerical solution in Fig.~\ref{fig:Z3stability}(d).

\section{Quadrupolar SPT} 
\label{sec:quad-spt}

The dipole symmetry considered in the previous section is the simplest type of modulated symmetry, where the modulation is linear in space. In this section we consider the next simplest case where the modulation is a quadratic function, corresponding to the conservation of quadrupole moment.

Recall that the conventional $\mathbb{Z}_N$ cluster model is constructed by dressing the operator $X_j$ with monopoles through the operator $Z^\dag_{j-1} \cdot Z_{j+1}$ or $Z_{j-1} \cdot Z^\dag_{j+1}$. Similarly, the $\mathbb{Z}_N$ dipolar SPT model is constructed by dressing $X_j$ by the dipolar domain-wall operator $Z_{j-1} Z^\dag_j  \cdot Z^\dag_j Z_{j+1}$. Following this guiding principle, an exactly soluble model with quadrupolar SPT order can be constructed as 
\begin{align}\label{eq:quadrupole_Hamiltonian}
    &H_{\rm Qu} =-\sum_{j} (a_{2j-1}+a_{2j}+h.c.) \nonumber
    \\
    &a_{2j-1}= Z_{2j-4} Z_{2j-2}^{-3} X_{2j-1} Z_{2j}^3 Z_{2j+2}^{-1} \nonumber
    \\
    &a_{2j} = Z_{2j-3}^{-1} Z_{2j-1}^3 X_{2j} Z_{2j+1}^{-3} Z_{2j+3} . 
\end{align}
(We use the Roman symbol `Qu' to refer to all things quadrupolar.) 
This a stabilizer Hamiltonian with two independent quadrupole symmetries acting on the odd and even sites
\begin{align}
{\rm Qu}_1 = \prod_j ( X_{2j-1} )^{j^2}~,\quad {\rm Qu}_2 = \prod_j ( X_{2j} )^{j^2}  .
\label{eq:Q1Q2-for-quadrupole} 
\end{align} 
The symmetry operators are indeed modulated quadratically with the spatial index $j$. Additionally, the model commutes with two monopole charge symmetries supported on the odd and the even sites separately, viz.\ the operators $Q_1$ and $Q_2$ given in \eqref{eq:charge-symmetry}. Finally, the model possesses both even- and odd-site dipole symmetries\footnote{Any translation-invariant model with quadrupole symmetry automatically also possesses dipole and monopole symmetries.}
\begin{align}
    D_1 = \prod_j ( X_{2j-1} )^{j}~, \quad D_2 = \prod_j ( X_{2j} )^{j}  .
\end{align}
We defined the stabilizers of $H_{\rm Qu}$ to be supported on seven sites, since this was the smallest support for which commuting stabilizers with the desired symmetry properties could be found. 

Let $\mathcal{T}$ be the operator which translates through one unit cell (two sites)), and we find the following algebra: 
\begin{align}
\mathcal{T}^{-1} {\rm Qu}_{1,2} \mathcal{T} & = Q_{1,2} D_{1,2}^2 {\rm Qu}_{1,2}  \nonumber \\ 
\mathcal{T}^{-1} {D}_{1,2} \mathcal{T} & = Q_{1,2} D_{1,2} . \label{eq:tranQuQD}
\end{align}
This algebra can be taken as the definition of a quadrupole symmetry. One can see that $Q_{1,2} D_{1,2}$ plays the role of a transition function connecting ${\rm Qu}_{1,2}$'s defined on overlapping segments of the lattice. The quadrupolar symmetry operators can thus also be understood as bundle symmetries, as discussed in Sec.~\ref{sec:periodic}.

The decorated domain wall picture applies nicely to the quadrupolar SPT model. For the odd-site-centered stabilizer $a_{2j-1}$ in \eqref{eq:quadrupole_Hamiltonian} we have the charge operator $X_{2j-1}$ dressed by a quadrupole-anti-quadrupole pair, written schematically as
\begin{align} [(1)_{2j-4} , (-2)_{2j-2} , (1)_{2j} ] , [(-1)_{2j-2} , (2)_{2j} , (-1)_{2j+2} ] ,
\end{align} 
where the numbers in parentheses mean the charges at a given site, and the subscripts are the coordinates. Accordingly, the first (second) bracket in the above represents a quadrupole (anti-quadrupole). For the even-site-centered stabilizer $a_{2j}$ the quadrupole and the anti-quadrupole positions are switched, 
\begin{align} [(-1)_{2j-3} , (2)_{2j-1} , (-1)_{2j+1} ] , [ (1)_{2j-1} , (-2)_{2j+1} , (1)_{2j+3} ] . \end{align} 
The decorated domain wall picture also explains why the quadrupolar SPT model is defined with a two-site unit cell, akin to the conventional $\mathbb{Z}_N$ cluster model, rather than a one-site unit cell. A fully translation-invariant model respecting a single $\mathbb{Z}_N$ quadrupole symmetry can be written down by invoking the domain wall picture, and leads to a 4-site Hamiltonian $H=-\sum_j(Z_{j-2} Z_{j-1}^{-3} X_{j} Z_{j}^3 Z_{j+1}^{-1}+h.c.)$. One can easily check, however, that the terms in the Hamiltonian do not commute with each other and thus do not assemble into a stabilizer Hamiltonian.

The ground state wave function for $H_{\rm Qu}$ on a periodic chain of even length $L$ is 
\begin{align}
|\Psi_{\rm Qu} \rangle & = {\cal U}_{\rm Qu} \left( \sum_{\mathbf{g}} |\mathbf{g} \rangle \right) \nonumber \\
{\cal U}_{{\rm Qu}} |\mathbf{g} \rangle & = \omega^{\sum_j g_{2j-1} ( 3g_{2j} - 3 g_{2j-2} + g_{2j-4} - g_{2j+2} )} |\mathbf{g} \rangle \nonumber \\ 
& = \omega^{\sum_j g_{2j} ( 3g_{2j-1} - 3 g_{2j+1} + g_{2j+3} - g_{2j-3})} |\mathbf{g} \rangle  . 
\end{align}
For completeness, we give the phase factor for the wave function in the case of an open chain of even length $L$,
\begin{align} 
& \omega^{g_{2} ( 3g_{1} - 3 g_{3} + g_{5})}\nonumber
\\
&\times\omega^{ g_{4} ( 3g_{3} - 3 g_{5} + g_{7} - g_{1}) +  \cdots + g_{L-4} ( 3g_{L-5} - 3 g_{L-3} +g_{L-1} - g_{L-7})} \nonumber \\ & \times \omega^{g_{L-2} ( 3g_{L-3} - 3 g_{L-1}  - g_{L-5})  + g_L (3 g_{L-1} - g_{L-3} ) }   \nonumber \\ 
& = \omega^{g_1 (3g_2 - g_4) + g_3 (3 g_4 - 3g_2 - g_6 ) + g_5 (3g_6 - 3g_4 + g_2 - g_8 ) + \cdots } \nonumber \\ & \times \omega^{g_{L-3} (3 g_{L-2} - 3g_{L-4} + g_{L-6} - g_L ) +  g_{L-1} (3g_L - 3g_{L-2} + g_{L-4} )}. 
\label{eq:quadrupolar-SPT-phase-factors} 
\end{align}
A similar expression can be found for a ground state on an open chain of odd length.

String order parameters of the quadrupolar SPT model \eqref{eq:quadrupole_Hamiltonian} are constructed according to the by-now familiar procedure. For the monopolar symmetries $Q_{1,2}$, we obtain
\begin{align} 
{\cal S}_{Q_1} &= a_{2j+1} \cdots a_{2j+2m-1} \nonumber \\ & = 
V_{2j}\left( \prod^m_{n=1}   X_{2j+2n-1} \right) 
V_{2j+2m}^{-1}\nonumber\\
{\cal S}_{Q_2} & = a_{2j} \cdots a_{2j+2m-2} \nonumber \\ & = 
V_{2j-1}^{-1}\left( \prod^m_{n=1}   X_{2j+2n-2} \right)  
V_{2j+2m-1}, 
\label{eq:quadrupole-string-order1}
\end{align}
where $V_{j}=Z_{j-2} Z_{j}^{-2} Z_{j+2}$. 
For the dipole and quadrupole symmetries, we find 
\begin{align}
{\cal S}_{D_1} &= a_{2j+1} a_{2j+3}^2 \cdots a_{2j+2m-1}^m \nonumber \\ & =  
W_{2j}\left( \prod^m_{n=1}   X_{2j+2n-1}^n \right)  W_{2j+2m}^{-1}
V_{2j+2m}^{-m}
, \nonumber\\
{\cal S}_{D_2} & = a_{2j} a_{2j+2}^2 \cdots a_{2j+2m-2}^m \nonumber \\ & = 
W_{2j-1}^{-1}
\left( \prod^m_{n=1}   X^n_{2j+2n-2} \right)  W_{2j+2m-1} 
V_{2j+2m-1}^m
, \nonumber 
\\
{\cal S}_{{\rm Qu}_1} &= a_{2j+1} a_{2j+3}^4 \cdots a_{2j+2m-1}^{m^2} \nonumber \\ & = 
\mathcal{V}_{2j}  \left( \prod^m_{n=1}   X^{n^2}_{2j+2n-1} \right)   \mathcal{V}_{2j+2m}^{-1}W_{2j+2m}^{-2m}V_{2j+2m}^{-m^2}, \nonumber\\
{\cal S}_{{\rm Qu}_2} & = a_{2j} a_{2j+2}^4 \cdots a_{2j+2m-2}^{m^2} \nonumber \\ & = \mathcal{V}_{2j-1}^{-1}  \left( \prod^m_{n=1}   X^{n^2}_{2j+2n-2} \right)   \mathcal{V}_{2j+2m-1}W_{2j+2m-1}^{2m}V_{2j+2m-1}^{m^2} ,
\label{eq:quadrupole-string-order2}
\end{align}
where $W_{j}=Z_{j-2} Z_{j}^{-1}$, ${\cal V}_{j}=Z_{j-2} Z_{j}$. 

From \eqref{eq:quadrupole-string-order1} and \eqref{eq:quadrupole-string-order2} one observes that the product of monopolar, dipolar, and quadrupolar operators (terms inside the parentheses) are flanked by quadrupolar, dipolar, and monopolar charges (terms on either side of the parentheses). This kind of structure is present in the conventional $\zn$ cluster model where, as in \eqref{eq:str-for-cluster}, the product of even(odd)-site symmetry operators are flanked by the charges at odd(even) sites. For the dipolar SPT model, the monopolar(dipolar) symmetry operators are flanked by dipoles(monopoles), as seen in \eqref{eq:string_ops}.  

We now place the Hamiltonian \eqref{eq:quadrupole_Hamiltonian} on an open chain and study the degenerate edge modes that arise due to the SPT order. Due to the extended nature of the stabilizers $a_{2j-1}, a_{2j}$, the model defined on an open chain of length $L$ has $a_1, a_2, a_3$ as well as $a_{L-2} , a_{L-1} , a_L$ missing. On such an open chain, we thus expect three degenerate $\zn$ modes on each edge. When acting on the ground state, these charges are equivalent to products of edge operators e.g. $Q_{1}|\Psi_{\rm{Qu}}\rangle=\mathcal{L}_{Q_{1}}\cdot\mathcal{R}_{Q_{1}}|\Psi_{\rm{Qu}}\rangle$. For even length $L=2\ell$, all the edge operators are listed as
\begin{alignat}{3}
    &\mathcal{L}_{Q_1}=V_{4}^{-1} X_1 X_3, ~~~~~~~~~~\, \mathcal{R}_{Q_1}=V_{L-2}X_{L-1},\nonumber
    \\
    &\mathcal{L}_{Q_2}=V_{3}X_2,~~~~~~~~~~~~~~~~~ \mathcal{R}_{Q_2}=V_{L-3}^{-1}X_{L-2}X_L,\nonumber
    \\
    &\mathcal{L}_{D_1}=W_{4}^{-1}V_{4}^{-2}X_1X_3^2
    ,~~~~ \mathcal{R}_{D_1}=W_{L-2}V_{L-2}^{\ell-1}X_{L-1}^{\ell}
    ,\nonumber
    \\
    &\mathcal{L}_{D_2}=W_{3} V_{3}X_2 , \nonumber 
    \\
    &\mathcal{R}_{D_2}=W_{L-3}^{-1}V_{L-3}^{-\ell+2}X_{L-2}^{\ell-1}X_L^{\ell}
    ,\nonumber
    \\
    & \mathcal{L}_{{\rm Qu}_1}=\mathcal{V}_4^{-1}W_4^{-4}V_{4}^{-4} X_1 X_3^4,\nonumber
    \\
    &\mathcal{R}_{{\rm Qu}_1}=\mathcal{V}_{L-2}W_{L-2}^{2\ell-2}V_{L-2}^{(\ell-1)^2}X_{L-1}^{\ell^2},\nonumber
    \\
    & \mathcal{L}_{{\rm Qu}_2}=\mathcal{V}_3W_3^2V_3X_2 ,\nonumber
    \\
    &\mathcal{R}_{{\rm Qu}_2}= \mathcal{V}_{L-3}^{-1}W_{L-3}^{-2\ell+4}V^{-(\ell-2)^2}_{L-3}X_{L-2}^{(\ell-1)^2}X_L^{\ell^2}.\nonumber
\end{alignat}
They obey the following algebra
\begin{alignat}{2}
&\mathcal{L}_{Q_1}\mathcal{L}_{\rm{Qu}_2}=\omega^{-2}\mathcal{L}_{\rm{Qu}_2}\mathcal{L}_{Q_1},~~~ &&\mathcal{R}_{Q_1}\mathcal{R}_{\rm{Qu}_2}=\omega^{2}\mathcal{R}_{\rm{Qu}_2}\mathcal{R}_{Q_1},\nonumber
    \\
&\mathcal{L}_{D_1}\mathcal{L}_{D_2}=\omega \mathcal{L}_{D_2}\mathcal{L}_{D_1}, ~~~&& \mathcal{R}_{D_1}\mathcal{R}_{D_2}=\omega^{-1} \mathcal{R}_{D_2}\mathcal{R}_{D_1},\nonumber\\
&\mathcal{L}_{\rm{Qu}_1}\mathcal{L}_{\rm{Qu}_2}=\omega^{-1}\mathcal{L}_{\rm{Qu}_2}\mathcal{L}_{\rm{Qu}_1},~~~&& \mathcal{R}_{\rm{Qu}_1}\mathcal{R}_{\rm{Qu}_2}=\omega^{-1}\mathcal{R}_{\rm{Qu}_2}\mathcal{R}_{\rm{Qu}_1},\nonumber
\end{alignat}
and
\begin{alignat}{2}
& \mathcal{L}_{\rm{Qu}_1}\mathcal{L}_{Q_2}=\omega^{-2}\mathcal{L}_{Q_2}\mathcal{L}_{\rm{Qu}_1},~~~&&
\mathcal{R}_{\rm{Qu}_1}\mathcal{R}_{Q_2}=\omega^2\mathcal{R}_{Q_2}\mathcal{R}_{\rm{Qu}_1},\nonumber\\
&\mathcal{L}_{\rm{Qu}_1}\mathcal{L}_{D_2}=\omega\mathcal{L}_{D_2}\mathcal{L}_{\rm{Qu}_1},~~~&&
\mathcal{R}_{\rm{Qu}_1}\mathcal{R}_{D_2}=\omega^{-1}\mathcal{R}_{D_2}\mathcal{R}_{\rm{Qu}_1},\nonumber \\
&\mathcal{L}_{D_1}\mathcal{L}_{\rm{Qu}_2}=\omega^{-1} 
    \mathcal{L}_{{\rm Qu}_2}\mathcal{L}_{D_1},~~~&& \mathcal{R}_{D_1}\mathcal{R}_{{\rm Qu}_2}=\omega \mathcal{R}_{{\rm Qu}_2}\mathcal{R}_{D_1},\nonumber
\end{alignat}
where trivial commutation relations have been omitted. We can pair up the left edge operators to form three independent Heisenberg algebras as 
\begin{align}
    \mathcal{L}_{D_1}\mathcal{L}_{D_2}&=\omega \mathcal{L}_{D_2}\mathcal{L}_{D_1},\nonumber
    \\
    \mathcal{L}_{Q_1}(\mathcal{L}_{\rm{Qu}_2}\mathcal{L}_{D_2})&=\omega^{-2}(\mathcal{L}_{\rm{Qu}_2}\mathcal{L}_{D_2})\mathcal{L}_{Q_1},\nonumber
    \\
    (\mathcal{L}_{\rm{Qu}_1}\mathcal{L}_{D_1}^{-1})\mathcal{L}_{Q_2}&=\omega^{-2}\mathcal{L}_{Q_2}(\mathcal{L}_{\rm{Qu}_1}\mathcal{L}_{D_1}^{-1}) . 
\end{align}
Similarly, the right edge operators can be paired up to form three independent Heisenberg algebras. These algebras enforce a minimal ground state degeneracy of $NK^2$ per edge, where $K=N/\text{gcd}(N,2)$.

In an apparent contraction, the Hamiltonian $H_{\rm  Qu}$ in \eqref{eq:quadrupole_Hamiltonian} has $N^3$ degenerate zero modes on each edge, which is equal to $NK^2$ only if $N$ is odd. This can be seen by calculating
${\cal U}_{\rm Qu}^\dag H_{\rm Qu} {\cal U}_{\rm Qu}$, which produces a trivial paramagnetic Hamiltonian which contains no terms acting within three lattice sites of each edge. 

The extra zero modes that appear in $H_{\rm Qu}$ when $N$ is even are however due to an accidental degeneracy, and are not topologically protected. To understand this, we first note that the algebra of operators acting on the space spanned by the $N^3$ zero modes of $H_\mcq$ is generated by the operators 
\begin{align}
    {\cal U}_{\rm Qu} X_1 {\cal U}_{\rm Qu}^\dag & =  X_1 Z_2^3 Z_4^{-1} \nonumber \\ 
    {\cal U}_{\rm Qu} X_2 {\cal U}_{\rm Qu}^\dag & = X_2 Z_1^3 Z_3^{-3} Z_5 \nonumber \\ 
    {\cal U}_{\rm Qu} X_3 {\cal U}_{\rm Qu}^\dag & =  X_3 Z_2^{-3} Z_4^3 Z_6^{-1}  
    \label{eq:quadrupolar-edge-op} 
\end{align}
together with the ${\cal U}_{\rm Qu} Z_j {\cal U}_{\rm Qu}^\dag =  Z_j$ for $j=1,2,3$, which can be derived using the phase factors in \eqref{eq:quadrupolar-SPT-phase-factors} for even $L$. The three operators obtained in \eqref{eq:quadrupolar-edge-op} mutually commute, and together with the $Z_j$ generate the three independent Heisenberg algebras which produce the $N^3$ degeneracy. 

Using the expressions for these operators, we see that this degeneracy can be reduced to the minimal value $NK^2$ by adding to $H_{\rm Qu}$ the following edge Hamiltonian: 
\begin{align}
    \Delta H=-(\mathcal{L}_{Q_1}^{K}+\mathcal{L}_{Q_2}^{K}+\mathcal{R}_{Q_1}^{K}+\mathcal{R}_{Q_2}^{K}+h.c.),
\end{align}
which commutes with $H_{\rm Qu}$  and all of the six symmetry generators. This lifting of the accidental edge degeneracy is similar to the discussions around \eqref{eq:edge_hamiltonian_dipole} for dipolar SPT.

In conclusion, we have constructed an exactly solvable quadrupolar SPT model, written down its ground states and string operators, and discussed its edge fractionalization. Construction of higher-order multipole SPT models following the strategy pursued here is also possible, and we leave a systematic investigation of their properties to future work. 

\section{Exponential SPT} 
\label{sec:6}
The dipole and quadrupole symmetries of the previous sections constitute some of the simplest examples of modulated symmetries~\cite{pollmann22}.
 In this section, we consider a more exotic case, where the modulation is by an {\it exponential} function of position.

To set the stage, we review the modified quantum clock model proposed in \cite{watanabe23}, which has Hamiltonian
\begin{align} H = - \sum_j ( Z^{-a}_j Z_{j+1}+h.c.) , \label{eq:watanabe-model} \end{align} 
with integer $a>1$ ($a=1$ being the ordinary $N$-state clock model). The relevant symmetry operator is
\be {\cal E} =\prod_j ( X_j )^{a^j},\ee 
which reduces to an ordinary monopolar charge operator $Q$ when $a =1\mod N$. For $a\mod N \neq 0,1$, each $X_j$ effectively creates a position-dependent charge $a^j$ that grows exponentially (mod $N$) with distance along the chain. Note that this model differs from the dipole and quadrupole models studied above in that it possesses only a single conservation law. 

Here we propose an SPT model protected by this kind of exponential symmetry. The commuting projector model we consider is 
\begin{align} H_{{\cal E}} & = - \sum_j (a_{2j-1} + a_{2j} + h.c.), \nonumber \\ 
a_{2j-1} & = Z^\dag_{2j-2} X_{2j-1} ( Z_{2j} )^a , \nonumber \\ 
a_{2j} & = ( Z_{2j-1} )^a X_{2j} Z^\dag_{2j+1}  . 
\label{espt}
\end{align} 
Note that $H_\mce$ reduces to the usual $\zn$ cluster Hamiltonian in (\ref{eq:cluster-H}) when $a = 1\mod N$, and shares the same two-site translational symmetry. The decorated domain wall picture still applies to the exponential SPT model above, with the understanding that the charges in the domain wall operators $Z^\dag_{2j-2} \cdot ( Z_{2j} )^a$ and $( Z_{2j-1} )^a \cdot Z^\dag_{2j+1}$ are exponentially modulated in space. 

On an open chain of length $L$, $H_\mce$ possesses two exponentially modulated symmetries, generated by the operators 
\begin{align} 
{\cal E}_1 =\prod_j ( X_{2j+1} )^{a^{j}} , ~~ {\cal E}_2=\prod_j ( X_{2j} )^{a^{L-j}} . 
\label{esym}
\end{align}
Unlike the dipolar SPT, when $a\mod N \neq 1,0$ (which we will specify to in what follows), the exponential SPT model in \eqref{espt} lacks a monopolar charge symmetry. 

The translation symmetry of the Hamiltonian $H_{\cal E}$ suggests that one can equally well shift $j \rightarrow j+j_0$ in the definition of the symmetry operators, \eqref{esym}. Indeed, such a shift results in 
\begin{align}
    {\cal E}_1 \rightarrow \left( {\cal E}_1 \right)^{a^{j_0}} ,  ~~~   {\cal E}_2 \rightarrow \left( {\cal E}_2 \right)^{a^{-j_0}} .
\end{align}
Here the equivalence relation holds among ${\cal E}_{1,2}$ raised to various powers, but does not involve any factor of charge or other multipoles. As a result, the model Hamiltonian $H_{\cal E}$ and its accompanying symmetry operators ${\cal E}_{1,2}$ are well-defined despite the absence of charge or other multipole conservation. 

For a periodic chain of length $L$, the exponential symmetry operators in \eqref{esym} are well-defined only if $a^L-1=0$ mod $N$. Following the discussion in Ref.~\cite{watanabe2022ground,watanabe23,delfino23}, we consider three scenarios:
\begin{itemize}[leftmargin=4mm]
    \item 
    $a$ and $N$ are coprime. In this case, from Euler's totient theorem, there always exists a finite integer $\varphi (N)$ such that $a^{\varphi (N)}-1=0$ mod $N$ with $\varphi (N)$ being the Euler's totient function. The exponential symmetry operator thus displays a periodicity under the translation by $\varphi (N)$ unit cells. When the system size satisfies $a^{L}-1=0$ ~mod $N$, the exponential symmetry is manifest under the periodic boundary condition, $X_{j+L} = X_j$.
    \item 
    $a= 0 ~\text{mod } {\rm rad}(N)$ where ${\rm rad}(N)$ is the radical of $N$ which is the product of the distinct prime numbers dividing $N$. In this case, there exists a finite integer $m$ such that $a^m = 0 ~\text{mod } N$ and the exponential symmetry operators in \eqref{esym} reduce to
    \begin{align} 
    {\cal E}_1 & =\prod_j ( X_{2j-1} )^{a^{j}} \rightarrow X^{a}_1 X^{a^2}_3  ...X^{a^{m-1}}_{2m-1},
    \end{align}
    by virtue of $( X )^{a^m}=1$ and all subsequent powers of $X$ being  1. The exponential symmetry thus becomes a local symmetry that only acts on a finite region of width $m$.\footnote{For local symmetry, we mean a symmetry whose generator has finite support. This is not to be confused with a gauge symmetry. } Our discussion on the exponential SPT phase is then not applicable to this case. 
    \item
    $a$ and $N$ are not coprime and $a\neq 0 ~\text{mod } {\rm rad}(N)$. 
    In this case, let us {\it define $N_a$ as the greatest divisor of $N$ that is coprime to $a$}. Then, when the exponential $\mathbb{Z}_N$ charge operator $\cal E$ is raised to $N_a^{\rm th}$ power, it becomes a local symmetry that acts only on a finite region. The exponential symmetry can then be viewed as a global symmetry that has infinite support extended by a local symmetry that has finite support. The exponential $\mathbb{Z}_N$ charge operator ${\cal E}$ does not exhibit any periodicity under lattice translations. Thus, the symmetry operator in \eqref{esym} cannot be defined properly for periodic boundary conditions of any system size $L$.  From Euler's theorem, there always exists a finite integer $\varphi (N_a)$ such that $a^{\varphi (N_a)}-1=0 \,\text{mod } N_a$ with $\varphi (n)$ being the Euler's totient function. Consequently, the exponential $\mathbb{Z}_{N_a}$ charge operator (which is a subgroup of the exponential $\mathbb{Z}_N$) still exhibits periodicity under the lattice translation.
\end{itemize}
Because of these subtleties, with periodic boundary conditions one needs to again resort to the concept of a bundle symmetry, as discussed in Sec.~\ref{sec:periodic}.

The ground state wave function of $H_{\cal E}$ is also a generalization of (\ref{eq:GS-cluster-H}) for the $\mathbb{Z}_N$ cluster model:
\begin{align}
|\Psi_{{\cal E}} \rangle & = {\cal U}_{{\cal E}} \left( \sum_{\mathbf{g}} |\mathbf{g}\rangle  \right) \nonumber \\ 
{\cal U}_{{\cal E}} |\mathbf{g}\rangle& =  \omega^{\sum_j g_{2j} ( a g_{2j-1}  - g_{2j+1} ) } |\mathbf{g}\rangle
.  \label{eq:GS-ecluster-H}
\end{align}
String-order operators characterizing the exponential SPT can be defined through the following products of stabilizers: 
\begin{align} 
{\cal S}_{{\cal E}_1} &= a_{2j+1} (a_{2j+3} )^a \cdots (a_{2j+2m-1} )^{a^{m-1}} \nonumber \\ & = Z^{\dag}_{2j} \left( \prod^m_{n=1}   ( X_{2j+2n-1} )^{a^{n-1}} \right)  ( Z_{2j+2m} )^{a^{m}}\nonumber\\
{\cal S}_{{\cal E}_2} & = ( a_{2j} )^{a^{m-1}} (a_{2j+2} )^{a^{m-2}}  \cdots a_{2j+2m-2}   \nonumber \\
&=Z_{2j-1}^{a^m} \left( \prod_{n=1}^m X_{2j+2n-2}^{a^{m-n}} \right) Z_{2j+2m-1}^{\dagger}. 
\label{str}
\end{align}
Physically, the string order parameters indicate that the clock patterns at two widely separated even (odd) sites are {\it locked} to the total exponential charge on the odd (even) sites in between.
Note also that as required, $\mcs_{\mce_{1,2}}$ reduce to the string order parameters of the cluster model \eqref{eq:str-for-cluster} when $a=1\mod N$. 
 
We can define fractionalized edge operators for a finite chain $1 \le j \le 2L$ as
\begin{align}
 &{\cal L}_1 \equiv X_1 Z_2^a, ~~ ~~~\,~{\cal L}_2 \equiv (Z_1)^{a^L} \nonumber \\ 
& {\cal R}_1 \equiv (Z_{2L})^{a^L} , ~~~ {\cal R}_2 \equiv  ( Z_{2L-1} )^a X_{2L}  , 
\end{align}
The exponential symmetries when acting on the ground states is equivalent to the products of these edge operators
\begin{align}
\mathcal{E}_1|\Psi_{{\cal E}} \rangle&={\cal L}_1\cdot {\cal R}_1^\dagger|\Psi_{{\cal E}} \rangle,\nonumber\\
\mathcal{E}_2|\Psi_{{\cal E}} \rangle&={\cal L}_2^\dagger\cdot {\cal R}_2|\Psi_{{\cal E}} \rangle.
\end{align}
On each edge, the symmetries are realized projectively, 
\begin{align} 
{\cal L}_1 {\cal L}_2 = \omega^{-{a^L} } {\cal L}_2 {\cal L}_1 , ~~ {\cal R}_1  {\cal R}_2 = \omega^{a^L} {\cal R}_2 {\cal R}_1 ,
\end{align}
with this relation reducing to (\ref{eq:LR-in-cluster-H}) for $a=1\mod N$.

As one can see from the algebra of the fractionalized edge operators, the way in which the projective symmetry is realized and consequently the degeneracy of the edge states depend crucially on the system size $L$, with the number of protected zero modes on each edge being $N/\gcd(N,a^L)$. When $a$ and $N$ are coprime, $\text{gcd}(N,a^L)=1$ for every $L$ and thus the edge degeneracy is always $N$. When $a$ and $N$ are not coprime and $a\neq 0 \text{ mod } {\rm rad}(N)$, the edge degeneracy is a non-increasing function of $L$ that equals $N_a$ for large $L$. The reason why the protected edge degeneracy varies with the system size is related to the fact that the exponential symmetries can be viewed as an extension of a global symmetry, that acts on the whole system, by a local symmetry, that acts on a finite region. This kind of ground state degeneracy depending on the system size is a manifestation of UV/IR mixing \cite{Seiberg:2020wsg,gorantla2021low} and has been documented in several lattice models in two dimensions~\cite{seiberg22a,seiberg23,oh22a,pace-wen,Gorantla:2022mrp,Gorantla:2022pii,delfino23,watanabe2022ground}. It is interesting that even an SPT phase can exhibit such kind of UV/IR mixing, as demonstrated by our example of exponential SPT. However, we emphasize that the exact ground state degeneracy is only the property of this stabilizer Hamiltonian. For a generic exponential SPT, the degeneracy can be lifted by finite size effects that decay with the system size. Therefore, only the edge degeneracy for large $L$, i.e.\ $N_a$, is truly robust against finite size effects.

One can perform a non-local duality to map the exponential SPT model in \eqref{espt} to a model similar to \eqref{eq:watanabe-model}\cite{watanabe23}: 
\begin{align} 
Z^\dag_{2j} & \rightarrow Z^\dag_{2j} \prod_{n=1} ( X_{2j+2n-1} )^{-a^{n-1}} \nonumber\\
Z^\dag_{2j+1} & \rightarrow Z^\dag_{2j+1} \prod_{n=1} ( X_{2j-2n+2} )^{-a^{n-1}} \nonumber\\
X_{j} & \rightarrow X_{j} . 
\label{dual}
\end{align}
Such a non-local transformation attaches an exponential charge string to the $Z$ operator without changing the Pauli algebra. The resultant dual Hamiltonian becomes
\begin{align} 
H= - \sum_j ( Z^{-1}_{2j-2} Z^{a}_{2j}+Z^{a}_{2j-1} Z_{2j+1}^{-1} +h.c. ) .
\label{esptdual}
\end{align}
This Hamiltonian is nothing but two copies of the model \eqref{eq:watanabe-model} studied in \cite{watanabe23} in the context of exponential symmetry breaking, defined on even and odd sites respectively. One can also view the model \eqref{esptdual} as the undecorated version of the exponential SPT model given in \eqref{espt}. 

The string order parameter of the exponential SPT phase in \eqref{str} becomes a two-point correlation function characterizing the long-range correlation of the exponential symmetry-breaking phase:
\begin{align} 
{\cal S}_1 & \rightarrow {\cal C}_1=Z^{\dag}_{2j} ( Z_{2j+2m} )^{a^{m}}, \nonumber \\ 
{\cal S}_2 & \rightarrow  {\cal C}_2= (Z_{2j-2m+1}^\dag )^{a^m} Z_{2j+1} . 
\end{align}
The operator mapping in \eqref{dual} embellishes each charge creation/annihilation operator at even (odd) sites with a string of exponential charges on odd (even) sites. Such embellishment exactly reproduces the decorated defect pattern of the SPT wave function. 

The non-local duality in \eqref{dual} establishes a connection between SPT and long-range order and generalizes the well-known mapping between the $\mathbb{Z}_N$ cluster model and the $N$-state quantum clock model~\cite{dhlee17} to $a>1$. This kind of non-local duality between SPT and symmetry-breaking states is rather universal. For instance, \cite{you2018subsystem,devakul2019fractal} showed that a subsystem symmetry-protected topological phase can be mapped to a subsystem symmetry-breaking state via a similar non-local duality. 

In summary, the exponential SPT model is a generalization of the $\mathbb{Z}_N$ cluster Hamiltonian with the exponentially modulated (instead of spatially uniform) charge domain walls. In this sense, the exponential model bears more resemblance to the charge cluster Hamiltonian than to the dipolar or quadrupolar SPT. 

\section{Summary and outlook}
\label{sec:7}

We have introduced a family of spatially modulated 1D symmetry-protected topological phases, with explicit examples protected by dipolar, quadrupolar, and exponentially modulated symmetries. For each symmetry, we constructed an exactly soluble lattice model and worked out the relevant diagnostics of SPT such as edge mode degeneracies and the symmetry fractionalization patterns. For the dipolar SPT, we performed a thorough analysis of the soluble model's MPS ground state, and demonstrated the inequivalence of the dipolar SPT to the more usual  SPT protected by monopolar symmetries. 

The defining characteristic of all our models is that the symmetries that protect their attendant SPT order are generated by operators that are spatially modulated. Understanding how these SPT orders behave under periodic boundary conditions led us to realize that the protecting symmetries are not always globally well-defined, prompting us to introduce the concept of bundle symmetry. Strictly speaking then, the phases considered in this work should thus be viewed as {\it bundle-symmetry protected topological phases.} In addition, modulated symmetries are distinguished by being generated by operators that do not commute with spatial translation. Consequently, spatial symmetry defects such as dislocations can permute different charge sectors, generating a fertile ground for exploring the effects of lattice defects.   

Many interesting questions remain open. For one, the nature of phase transitions of modulated SPT states into trivial states and the accompanying critical theories are likely to differ from those of well-known SPT to non-SPT transitions studied in the past~\cite{dhlee17,verresen2017,prembabu22}. Another fruitful and related line of inquiry is the investigation of the relationship between modulated SPT and symmetry-breaking phases. Although the mapping of the exponential SPT model to the symmetry-breaking model of the same symmetry can be readily worked out (see Sec. \ref{sec:6}), the analogous mappings for the dipolar and quadrupolar models are less clear.

It is known that the $\mathbb{Z}_2 \times \mathbb{Z}_2$ monopolar SPT symmetries have a well-known physical realization in the spin-1 Haldane chain. It is then interesting to ask whether one can construct a realistic spin model corresponding to, for example, $\mathbb{Z}_N \times \mathbb{Z}_N$ dipolar SPT phase and explore its physical consequences. 

Lastly, our study of modulated SPT phases has led to a generalization of global symmetries: we found that bundle symmetries are responsible for the protection of this topological order. It would be fascinating to explore this novel concept in a broad ranger of circumstances, such as higher dimensions, spontaneous symmetry breaking, gauging, and beyond.

\acknowledgments We are grateful to Meng Cheng, Johannes Feldmeier, Hyunyong Lee, Jongyeon Lee, Rahul Sahay, David T. Stephen and Haruki Watanabe for helpful discussions and feedback. DMRG simulations were performed using the TeNPy Library \cite{tenpy}, which was inspired by a previous library \cite{Kjaell13}. J.H.H. was supported by the National Research Foundation of Korea(NRF) grant funded by the Korea government(MSIT) (No. 2023R1A2C1002644). He also acknowledges financial support from EPIQS Moore theory centers at MIT and Harvard, where this work was initiated. H.T.L. is supported in part by a Croucher fellowship from the Croucher Foundation, the Packard Foundation and the Center for Theoretical Physics at MIT. He also thanks the ``Paths to Quantum Field Theory 2023" workshop for hospitality during the course of this project. R.V. is supported by the Simons Collaboration on Ultra-Quantum Matter, which is a grant from the Simons Foundation (618615, Ashvin Vishwanath). This work was completed in part at Aspen Center for Physics (RV, JHH, YY), which is supported by National Science Foundation grant PHY-2210452 and Durand Fund. The authors have been listed in alphabetical order. 

\appendix
\section{Parafermion formulation of dipolar SPT}\label{app:para}
In this appendix, we present the parafermion mapping of the dipolar SPT model. Generalizing the Jordan-Wigner transformation~\cite{fendley2014free}, one can map our $\mathbb{Z}_N$ dipolar SPT model to an interacting parafermion chain whose boundary parafermion mode plays the role of the edge zero modes in the dipolar SPT phase.

Recall that the dipolar SPT Hamiltonian \eqref{eq:H-dipolar} is given as a sum of stabilizers $H_{D} =-\sum_j ( a_j + a_j^\dag )$ where $a_j = Z_{j-1} Z^\dag_j X_j Z^\dag_j Z_{j+1}$. We can make a parafermion representation of $H_D$ by introducing a pair of parafermion operators
\begin{align}
    \chi_i   = Z_i \left( \prod_{j<i} X_j \right) , ~\psi_i = Z_i X_i \left( \prod_{j < i} X_j \right) . 
\end{align}
The $\chi_i,\psi_i$ are the parafermion operators satisfy $(\chi_i )^N = 1$, $(\psi_i )^N = 1$, $\chi \chi^\dag = \psi \psi^\dag =1$, and their commutation relations are
 \begin{align}
    \psi_i \psi_j = 
    \begin{cases}
    \omega^{-1} \psi_j \psi_i,~~ &i>j
    \\
    \omega \psi_j \psi_i,~~ &i<j
    \end{cases},\nonumber
    \\
    \chi_i \chi_j = 
    \begin{cases}
    \omega^{-1} \chi_j \chi_i,~~ &i>j
    \\
    \omega \chi_j \chi_i,~~ &i<j
    \end{cases},\nonumber
    \\
    \chi_i \psi_j=
    \begin{cases}
    \omega^{-1} \psi_j \chi_i,~~ &i>j
    \\
    \omega \psi_j \chi_i,~~ &i\leq j
    \end{cases}
\end{align}
Inversely, one can map
 \begin{align}
     Z_{i-1} Z^\dag_{i} = \psi_{i-1} \chi^\dag_{i} , ~~     Z^\dag_{i} Z_{i+1} = \chi_{i+1} \psi^\dag_{i} , ~~  X_i = \chi^\dag_i \psi_i . 
 \end{align}
 
The dipolar SPT Hamiltonian can thus be interpreted as a parafermion chain,
\begin{align}
    H& = - \sum_j \left[ (Z_{j-1}Z^\dag_{j}) X_j (Z^\dag_{j} Z_{j+1})  + h.c. \right] \nonumber\\ 
     &\rightarrow - \sum_j \left[ \psi_{j-1} ( \chi^\dag_j )^2 \psi_j \chi_{j+1} \psi^\dag_i + h.c. \right]  \nonumber \\ 
     & = - \sum_j \left[  \omega \psi_{j-1} ( \chi^\dag_j )^2 \chi_{j+1} + h.c. \right].  
\end{align}
After some algebra, we find that there exist two sets of parafermion operators that commute with the Hamiltonian. 
\begin{alignat}{2}
& L_1=\chi_1,~~~ &&R_1=\chi_L \prod_{j<L} (\psi^\dag_{j}\chi_j),\nonumber\\
&
L_2=\chi_2,~~~ &&R_2=\chi_{L-1} \prod_{j<L-1} (\psi^\dag_{j}\chi_j )\chi^\dag_L\psi_{L}
\end{alignat}
These operators exactly match the edge operators of the dipolar SPT chain in \eqref{eq:dipolar-edge-1}.

\section{Precise definition of bundle symmetries} \label{app:symm_bundles}

This appendix is devoted to formulating a mathematically precise definition of the {\it bundle symmetries} introduced in Sec.~\ref{sec:periodic} of the main text. 

Consider a local Hamiltonian $H$ which acts on a collection of $d$-dimensional qudits placed on the vertices of a $D$-dimensional spatial lattice $M$. Let $\{ A_\a\} $ denote a collection of contractible\footnote{Since our Hamiltonians are defined on lattices, this use of the word ``contractible'' is a bit colloquial. More precisely, we will actually think of ourselves as working on a CW complex, the 0-cells of which constitute the lattice $M$. We then call a collection of points in $M$ {\it contractible} if the $D$-manifold defined by the union of all $D$-cells whose boundary 0-cells are all contained in $M$ is contractible. } subregions which together provide an open cover for $M$ (so that all of the intersections $A_\a \cap A_\b$ are themselves contractible). 
A {\it bundle symmetry} $E_G$ over $M$ is defined by a projection $\pi : E_G \ra M$ and a certain collection of operators defined on the $A_\a$ which we will define momentarily. On each patch $A_\a$, these operators are defined using local inverses of $\pi$, which identify the preimage of $A_\a$ under $\pi$ with the product space $ A_\a\times F$, where the fiber $F=U(d)$ is the $d$-dimensional unitary group ($d$ being the local Hilbert space dimension). We will write these local inverses as $\pi_g\inv$, where for now $g$ is a formal symbol indexing the different inverses. For each $\pi_g\inv$ and each patch $A_\a$, we define the {\it symmetry section} $\vp_{g\a}$ as 
\begin{equation} \vp_{g\a} \equiv \prod_{j\in A_\a} \pi_g\inv(j)_{F},\end{equation}
 where $\pi_g\inv(j)_{F}$ denotes the restriction of $\pi_g\inv(j)$ to the fiber $F=U(d)$,
 and where for all $A_\a$ we require that $\vp_{g\a}$ behave like an internal symmetry of $H$ on the interior of $A_\a$, meaning that $H\vert_{A_\a}$, the restriction of the Hamiltonian $H$ to $A_\a$, commutes with $\vp_{g\a}$
 \be
 \label{almostsym}
 [H\vert_{A_\a},\vp_{g\a} ]=0~.
 \ee
 Note that the set of operators satisfying \eqref{almostsym} are closed under matrix multiplication and taking inverses. The $\vp_{g\a}$ thus form a group, explaining our choice to index them by the symbol $g$ (and allowing us to write e.g. $\vp_{g\a} \vp_{h\a} \equiv \vp_{gh\a}$ and $\vp_{g\inv\a} \equiv \vp_{g\a}^\da$). Finally, for each pair of patches $A_\a ,A_\b$ with $A_\a \cap A_\b \neq \emptyset$, we define the transition operators
 \begin{align} t_{g\a,h\b} \equiv (\vp_{g\a}^\da \vp_{h\b})|_{A_\a\cap A_\b}.\end{align} 
Note that $t_{g\a,h\b}$ is itself a symmetry section on $A_\a \cap A_\b$. 
Also note that on triple overlaps of subsystems $A_\a \cap A_\b \cap A_\g \neq \emptyset$, we have a ``cocycle condition'' $t_{g\a,h\b} t_{h\b,k\g} t_{k\g,g\a} = \unit.$

 In the language of bundle symmetries, an ordinary internal global symmetry is simply a global section of $E_G$, viz.\ a symmetry section $\vp_g$ which admits an extension of its range from the collection of patches $A_\a$ to the entire spatial manifold $M$, with $[\vp_g(M), H] = 0$. More precisely, a global symmetry is a symmetry section characterized by the property that $t_{g\a,g\b} = \unit$ for all patches $A_\a, A_\b$ with nonzero intersection. For sections $\vp_g$ which do not admit an extension to $M$, the associated ``symmetry'' can only be defined patch-by-patch, as was shown to be the case for the $\zn$ dipole symmetry considered above. If choosing $t_{g\a,g\b} = \unit$ is obstructed for any $g$, we will call the bundle symmetry ``nontrivial''.

One-dimensional systems with spatially modulated symmetries \cite{sala2022dynamics} provide the simplest examples of nontrivial bundle symmetries. When the spatial manifold is a line, the bundle symmetries that arise are always trivial, since a single patch $A_\a$ suffices to cover all of space. A circle requires at least three patches however, and nontrivial bundles are possible. Consider a translation-invariant system with a local symmetry section $\vp_{g\a} = \prod_{j\in A_\a} U_j^{g(j)}$, where $U_j$ is a given unitary and $g(j)$ is some function of the spatial coordinate. Translation invariance means that different sections $\vp_{g\a}$ can be chosen by replacing $g(j)$ with $g(j+j_\a)$ for any integer $j_\a$; this gives transition operators of the form $t_{g\a,g\b} = \prod_{j \in A_\a \cap A_\b} U_j^{g(j+j_\a) - g(j+j_\b)}$.

\bibliography{ref}

\end{document}